\documentstyle[epsf,eqsecnum,floats,preprint,aps]{revtex}

\tighten

\input epsf
\begin{document}
\newcommand{\beq}{\begin{equation}}
\newcommand{\eeq}{\end{equation}}

\newcommand{\be}{\begin{equation}}
\newcommand{\ee}{\end{equation}}
\newcommand{\bea}{\begin{eqnarray}}
\newcommand{\eea}{\end{eqnarray}}
\newcommand{\PSbox}[3]{\mbox{\rule{0in}{#3}\includegraphics{#1}\hspace{#2}}}
\overfullrule=0pt
\def\Int{\int_{r_H}^\infty}
\def\d{\partial}
\def\e{\epsilon}
\def\M{{\cal M}}
\def\high{\vphantom{\Biggl(}\displaystyle}
\catcode`@=11
\def\@versim#1#2{\lower.7\p@\vbox{\baselineskip\z@skip\lineskip-.5\p@
    \ialign{$\m@th#1\hfil##\hfil$\crcr#2\crcr\sim\crcr}}}
\def\simge{\mathrel{\mathpalette\@versim>}} %
\def\simle{\mathrel{\mathpalette\@versim<}} %
\catcode`@=12 

\def\pr#1#2#3#4{Phys. Rev. D {\bf #1}, #2 (19#3#4)}
\def\prl#1#2#3#4{Phys. Rev. Lett. {\bf #1}, #2 (19#3#4)}
\def\prold#1#2#3#4{Phys. Rev. {\bf #1}, #2 (19#3#4)}
\def\np#1#2#3#4{Nucl. Phys. {\bf B#1}, #2 (19#3#4)}
\def\pl#1#2#3#4{Phys. Lett. {\bf #1B}, #2 (19#3#4)}

\rightline{CU-TP-940}
\rightline{hep-th/9905223}
\vskip 1cm

\begin{center}
\ \\
\large{{\bf Magnetic monopoles near the black hole threshold}}
\ \\
\ \\
\ \\
\normalsize{Arthur Lue\footnote{\tt lue@phys.columbia.edu} and 
Erick J. Weinberg\footnote{\tt ejw@phys.columbia.edu}}
\ \\
\ \\
\small{\em Department of Physics \\
Columbia University \\
New York, NY 10027}

\end{center}

\begin{abstract}

We present new analytic and numerical results for self-gravitating
SU(2)-Higgs magnetic monopoles approaching the black hole threshold.
Our investigation extends to large Higgs self-coupling, $\lambda$, a
regime heretofore unexplored.  When $\lambda$ is small, the critical
solution where a horizon first appears is extremal Reissner-Nordstrom
outside the horizon but has a nonsingular interior.  When $\lambda$ is
large, the critical solution is an extremal black hole with
non-Abelian hair and a mass less than the extremal Reissner-Nordstrom
value.  The transition between these two regimes is reminiscent of a
first-order phase transition.  We analyze in detail the approach to
these critical solutions as the Higgs expectation value is varied, and
compare this analysis with the numerical results.

\end{abstract}

\setcounter{page}{0}
\thispagestyle{empty}
\maketitle

\eject

\vfill

\baselineskip 16pt plus 2pt minus 2pt

\section{Introduction}

The physics of a nonsingular spacetime is qualitatively distinct from
a that of a spacetime exhibiting a black hole.  However, families of
spacetimes exist that may be viewed as interpolating between the two.
These spacetimes are nonsingular and have no horizons;  nevertheless,
they have a region whose metric can be made arbitrarily close to that
of the exterior region of a black hole.  In the limiting case, the
inner boundary of this region takes on the characteristics of an
extremal horizon, even though no curvature singularity develops in the
interior.  By studying such solutions, one may gain further insight
into the properties of black holes.

One approach to the construction of such ``almost black holes'' begins
with a spontaneously broken Yang-Mills theory that has magnetic
monopole solutions.  For small Higgs expectation values, these
monopole solutions persist when gravitational effects are included.
However, as the Higgs expectation value is increased toward a critical
value on the order of the Planck mass, the monopole solutions begin to
approximate black holes with finite mass and nonzero horizon radius.

Such gravitating monopoles, as well as the related magnetically
charged black holes with hair, have been studied previously
\cite{orig1,ortiz,orig2,orig2b,AB}.  For a review article and
recent work in related subjects, see \cite{review}.
In this paper we investigate these solutions in more detail,
concentrating on some aspects that were not previously noted.  Most
notably, we find that two distinct types of behavior, with
qualitatively different extremal black hole limits, can occur.  Our
focus here is on the detailed properties of these solutions and their
behavior as they approach the black hole limit.  We describe elsewhere
\cite{future} how these solutions can be used as approximate black holes
that provide insight into the transition from a nonsingular spacetime
to one with horizons.

We work in the context of an SU(2) gauge theory that is spontaneously
broken to U(1) by a triplet Higgs field $\phi$ with vacuum expectation
value $\langle \phi \rangle = v$.  The elementary particle spectrum of
this theory contains a pair of vector mesons $W^\pm$ that carry U(1)
electric charge and have mass $ev$ (where $e$ is the gauge coupling),
as well as a massless photon and a massive electrically neutral Higgs
particle.  In flat space the classical field equations have a
spherically symmetric monopole solution with U(1) magnetic charge
$4\pi/e$ and a mass of order $4\pi v/e$.  This has a central ``core''
region, of radius $R_{\rm core} \sim 1/m_W$, in which there are
nontrivial matter fields.  Beyond this radius is a ``Coulomb'' region
in which the massive fields fall off exponentially fast, leaving only
a long-range Coulomb magnetic field.

In studying the behavior of these solutions in the presence of
gravity, we assume spherical
symmetry, and so can write the metric in the form
\begin{equation}
  ds^2 = B(r)dt^2 - A(r)dr^2 - r^2 (d\theta^2 + \sin^2 \theta d\phi^2)
  \ .
\label{metricansatz}
\end{equation}
It is often convenient to rewrite $A$ in terms of a mass function
$m(r)$ defined by
\begin{equation}
    {1\over A(r)} = 1 - {2G m(r) \over r}
\label{massdef}
\end{equation}
with $m(\infty) \equiv M$.  For a configuration to be nonsingular at
the origin, $A(0)=1$ and $m(0)=0$.  A horizon occurs when $1/A$ has a
zero or, equivalently, when $m(r)/r = 1/2G$.

A benchmark with which to compare our results is provided by the
Reissner-Nordstrom metric, with
\begin{equation}
      B(r) ={1\over A} = 1 -{2MG\over r} + {Q^2 G\over 4\pi r^2}\ .
\end{equation}
If $M > \sqrt{Q^2/4\pi G}$, this describes a black hole solution with
a charge (either magnetic or electric) $Q$ and an outer horizon
determined by the larger of the two zeroes of $1/A$.  If instead $M <
\sqrt{Q^2/4\pi G} $, there is no horizon separating the curvature
singularity at $r=0$ from the asymptotic regions and a naked
singularity results.  The boundary between these two regimes, $M=
\sqrt{Q^2/4\pi G}$, gives the extremal Reissner-Nordstrom black hole.
For $Q= 4\pi/e$, the case with which we will be concerned in this
paper, the extremal black hole has a horizon radius
\begin{equation}
    r_0 = \sqrt{4 \pi G \over e^2}
\end{equation}
and a mass 
\begin{equation}
    M_0 = \sqrt{4 \pi  \over G e^2}\ .
\end{equation}

When $v \ll M_{\rm Pl}$ the gravitational effects on the monopole are
relatively small.  The metric at large distances approaches the
Reissner-Nordstrom form, with $\sqrt{Q^2/4\pi G} \gg M$.  There is no
singularity, because the actual metric deviates from the
Reissner-Nordstrom form when $r \simle R_{\rm core}$.  One finds that
$1/A = 1$ at the origin, decreases to a minimum at a radius of order
$R_{\rm core}$, and then increases monotonically.  As $v$ is
increased, the core shrinks and the minimum of $1/A$ moves inward and
becomes deeper.  In most cases, this continues until $1/A$ develops a
double zero, corresponding to an extremal horizon, when $v=v_{\rm cr}$.
A slightly different behavior is found for very small Higgs
self-coupling \cite{orig2}.  In this case, the solution varies continuously
as $v$ is increased up to a value $v_{\rm max}$.  Although the minimum
value of $1/A$ is still nonzero at this point, static solutions do not
exist for higher values of $v$.  Instead, these solutions join
smoothly on to a second branch of solutions for which $(1/A)_{\rm
min}$ decreases as $v$ is {\it decreased} from $v_{\rm max}$ to a
critical value $v_{\rm cr}$, where the extremal horizon develops.

For small values of the Higgs mass, the extremal horizon of the
critical solution occurs at the Reissner-Nordstrom radius $r_0$, in
the exterior Coulomb region of the monopole solution.  The matter
fields take on their vacuum value everywhere outside the horizon and
the exterior metric is exactly that of an extremal Reissner-Nordstrom
black hole.  When the Higgs to vector mass ratio is greater than about
12, a regime not explored in depth in previous studies, the behavior
is rather different.  The horizon in the monopole core at a radius
$r_* < r_0$ that decreases with increasing Higgs mass.  In this case
there are nontrivial matter fields, or ``hair'', outside the horizon.
The transition between the two regimes is not smooth, but instead is
reminiscent of a first order transition.

The remainder of this paper is organized as follows.  In Sec.~II we
outline the general formalism and define our conventions.  In Sec.~III
we use numerical methods to obtain monopole solutions to the field
equations.  We describe in detail their behavior as $v$ is increased
towards its critical value.  The critical solutions that are the
limits of these families of monopole solutions are characterized by
the presence of extremal horizons.  In Sec.~IV, we use analytic
methods to study the properties of these critical solutions, focusing
on the behavior near the horizon.  We show that the problem of finding
a solution with an extremal horizon can be formulated as a pair of
boundary value problems, one for the region $0<r<r_*$ and one for
$r_*<r<\infty$, that must be solved simultaneously.  The conditions
for a solution to these places strong constraints on the behavior of
the fields near the horizon.  These constraints allow only two types
of behavior, one associated with a core region horizon, the other with
a Coulomb region horizon, which we examine in detail in Sec.~V.  In
Sec.~VI we compare these analytic predictions with our numerical
results.  Section~VII contains some concluding remarks.  The Appendix
contains details of the numerical investigation of the transition
region between the two types of critical solutions.

\section{General Formalism}

We consider an SU(2) gauge theory with a triplet Higgs field
$\phi^a$ whose self-interactions are governed by the scalar field
potential 
\begin{equation}
V(\phi) = {\lambda \over 2}\left(\phi^a\phi^a -v^2\right)^2 \ .
\end{equation} 
In flat spacetime, this theory has nonsingular monopole solutions,
with magnetic charge $4\pi/e$, that are described by the spherically
symmetric ansatz
\begin{eqnarray}
\phi^a &=& v {\hat r}^a h(r)
\label{phiansatz}	\\
A_{ia} &=& \epsilon_{iak} {\hat r}^k\, {1 -u(r) \over er}
\label{Aansatz}		\\
A_{0a} &=& 0 \ .
\label{A0}
\end{eqnarray}
While it is clear that $vh(r)$ is the magnitude of the
Higgs field, the meaning of $u(r)$ is somewhat obscured by this
``radial gauge'' ansatz.  By applying a singular gauge transformation
that makes the direction of the Higgs field uniform, one finds that
$u(r)/er$ is equal to the magnitude of the massive vector field, and
so it, like $1-h(r)$, should be expected to vanish exponentially fast outside
the monopole core.

The generalization of this ansatz to curved spacetime is
straightforward.  Since we are considering only static spherically
symmetric solutions, we can take the metric to be of the form of
Eq.~(\ref{metricansatz}).
The matter part of the action is
\begin{equation} 
   S_{\rm matter} = -4\pi \int dt\, dr\, r^2 \sqrt{AB} 
      \left[ {K(u,h)\over A} + U(u,h) \right]
\label{matteraction}
\end{equation}
where 
\begin{eqnarray}
K &=& \frac{1}{e^2r^2}\left(\frac{du}{dr}\right)^2 
	+ \frac{v^2}{2}\left(\frac{dh}{dr}\right)^2
\nonumber	\\
U &=& {(u^2-1)^2\over 2e^2r^4} + {v^2u^2h^2\over r^2} 
    + {\lambda v^4\over 2} (h^2-1)^2 \ .		\nonumber
\end{eqnarray}

One may view $U(u,h)$ as a position-dependent potential.  It has
several stationary points, which we enumerate here for later
reference: 

\begin{enumerate}
	\item $u = 0 $, $h=\pm 1$.  This is a local minimum of $U$
	for $r > 1/ev $.
	\item $u= \hat u(r)$, $h=\hat h(r)$, or $u= -\hat u(r)$,
	$h=-\hat h(r)$, where
		\begin{eqnarray}
		\hat u &=& \sqrt{\frac{\lambda(1-e^2v^2r^2)}{\lambda-e^2}}
     		\label{uhateq} \\ \hat h &=& \sqrt{\frac{\lambda -
     		e^2/(evr)^2}{\lambda-e^2}}\ .  \label{hhat}
		\end{eqnarray}
	These are real only for $r$ lying between $1/ev$ and $1/\sqrt{\lambda}
	\, v$.  In that range it is a minimum of $U$ if $\lambda/e^2 >1$, but
	a saddle point if $\lambda/e^2 <1$.  If $\lambda/e^2=1$, this solution
	is replaced by a degenerate set of local minima, with $h^2 + u^2 =1$,
	that exist only for $evr =1$.
	\item $ u =h = 0$.  This is never a local minimum of $U$.
	\item $u = \pm 1$, $h=0$.  This is a local minimum of $U$
	for $r < 1/\sqrt{\lambda} \, v$.
\end{enumerate}
We will see that these stationary points are key to understanding the
local existence of extremal horizons in monopole systems.

For static solutions, the matter fields obey the equations
\begin{eqnarray}
{1\over\sqrt{AB}}\ \frac{d\ }{dr}\left(\frac{\sqrt{AB}}{A}\frac{du}{dr}\right)
    &=& \frac{e^2r^2}{2}\frac{\d U}{\d u} = {u(u^2-1)\over r^2} + e^2v^2 uh^2
\label{ueqn}	\\
{1\over r^2\sqrt{AB}}\ \frac{d\ }{dr}
\left(\frac{r^2\sqrt{AB}}{A}\frac{dh}{dr}\right)
    &=& {1 \over v^2}
	\frac{\d U}{\d h} = {2hu^2\over r^2} + 2\lambda v^2 h(h^2 -1)\ .
\label{heqn}
\end{eqnarray}
These must be supplemented by the two gravitational field equations
\begin{eqnarray}
\frac{1}{\sqrt{AB}}\frac{d\sqrt{AB}}{dr} &=& 8\pi G r K
\label{ABeqn}	\\
r\ \frac{d\ }{dr}\left(\frac{1}{A}\right) &=& \left(1 - 8\pi Gr^2U\right)
				- \frac{1}{A}\left(1 + 8\pi Gr^2K\right)\ .
\label{Meqn}
\end{eqnarray}
Note that, up to a rescaling of distances, the solutions of these
equations depend only on the dimensionless parameters\footnote{These
parameters are related to those in Ref.~\cite{orig1} by $\mu =a$,  
in Ref.~\cite{ortiz} by $\alpha = 2{{b}}$ and $\beta = {{a}}$, and in 
Ref.~\cite{orig2} by
$\alpha = \sqrt{{a}/2}$ and $\beta = 2\sqrt{{b}}$.}
${{a}} = 8\pi Gv^2$ and ${{b}} = \lambda/e^2 = (m_H/2 m_W)^2$.

Integration of Eq.~(\ref{ABeqn}) gives $B(r)$ in terms of the
remaining functions.  With the boundary condition $A(\infty)B(\infty)=1$,
corresponding to the conventional normalization of $t$, we have
\begin{equation}
   B(r) = {1 \over A(r)} \exp\left[-16\pi G\int_r^\infty dr'\, r'
   K\right]\ .
\label{Bsolution}
\end{equation}
Using this result to eliminate $B(r)$ from the remaining field
equations leaves one first-order and two second-order equations to be
solved.  A solution of these is determined by five boundary
conditions.  Requiring that the fields be nonsingular at the origin
gives three of these, $u(0)=1$, $h(0)=0$, and $A(0)=1$.  Two more,
$u(\infty)=0$ and $h(\infty)=1$, follow from the finiteness of the
energy.  Additional boundary conditions arise when a horizon is
present.  While these are not relevant for our numerical solutions,
which are all regular monopoles, they play an important
role in our analysis of the extremal black hole configurations that
are the limiting points of sequences of monopole solutions.

\section{Monopole Solutions}

\subsection{The method}

In our search for regular solutions and their approach to criticality,
let us identify an energy functional on the space of static configurations.
Following van Nieuwenhuizen, Wilkinson, and Perry \cite{vanN}, one can
write down an action which is a functional of just $u(r)$ and $h(r)$
by not only eliminating $\sqrt{AB}$ using Eq.~(\ref{ABeqn}), but also
eliminating $A$ by integrating Eq.~(\ref{Meqn}) subject to the
boundary condition $A(0)=1$.  The complete action (i.e., gravitational
plus matter) then becomes
\begin{eqnarray}
	S &=& - \int dt\ E
\nonumber	\\
	E &=& 4\pi\int_0^\infty dr\, \left\{
		 \frac{r}{8\pi G}\frac{d\ }{dr}\sqrt{AB}
	\left[1-\frac{1}{A}\right]\right.
\label{action}	\\
&&\ \ \ \ \ \ \ \ \ \ \ \ \ \ \ \ \ \ \ \ \ \ \ \left. +\ r^2\sqrt{AB}
	     \left[ {K(u,h) \over A} + U(u,h) \right]\right\}\ .
\nonumber
\end{eqnarray}
with $\sqrt{AB}$ and $A$ understood to be
implicit functionals of $u(r)$ and $h(r)$.   Configurations that extremize this
action are solutions to Eqs.~(\ref{ueqn})-(\ref{Meqn}), with the energy $E$
being equal to the mass $M= m(\infty)$ \cite{vanN}.  
This energy functional is not bounded
from below, even when satisfying the appropriate boundary conditions,
unless the $1/A(r)$ that corresponds to a given matter field
configuration $\{u(r),h(r)\}$ is everywhere greater than zero.  This
restriction is not serious for our purposes, since we are only
interested in static regular monopole solutions and their approach to
criticality; all such solutions satisfy the condition that $1/A(r)>0$
everywhere.

Thus, we find {\em regular} solutions to the field equations by
seeking extrema of this energy functional that satisfy the boundary
conditions at $r=0$ and $r= \infty$.  This is equivalent to solving
$\delta E/\delta u =  \delta E/\delta h = 0$.  We do this
numerically by solving the alternative system of equations
\begin{eqnarray}
	\frac{d^2u}{dt^2} + \Gamma \frac{du}{dt} = \frac{\delta E}{\delta u}
\nonumber 	\\
	\frac{d^2h}{dt^2} + \Gamma \frac{dh}{dt} = \frac{\delta E}{\delta h}
\label{numerical}
\end{eqnarray}
for an appropriate damping factor $\Gamma$ and taking the final steady
state solution as the result.   This approach
is analogous to having a massive particle roll on a
manifold determined by $E$ while the particle's motion is viscously
damped, eventually coming to rest at some local minimum of the energy.

The discretization used to numerically implement this algorithm places
a limit on how close we can approach the critical solutions in which
the extremal horizon has actually formed.  The errors in the fields
are proportional to the square of the spatial step size.  As a result
is that we can only obtain solutions in which the minimum
value of $1/A$ is at least ${\cal O}[(\Delta r)^2]$.

One should contrast our relaxation method with the algorithms used
in previous analyses \cite{orig1,ortiz,orig2} that employ shooting from the
origin.  In these, a choice is made for the values of the fields  
and their first spatial derivatives at the origin, and the equations of motion
are then used to integrate out towards infinity.  The initial choice is then
adjusted to ensure that the boundary conditions at infinity are
satisfied.  With this approach, difficulties appear with large
${{b}}$ because of the extreme sensitivity at the origin to small
perturbations.  No such problem exists with relaxation, and as a
result we have been able to obtain solutions for large ${{b}}$.  
However, although solutions that are unstable under perturbations may
be found by shooting, such solutions cannot be obtained by our
relaxation method.

In particular, previous work \cite{orig2} has shown that for ${{b}}
\simle 0.1$ monopole solutions exist for a range of values of ${{a}}$
that are greater than the value ${{a}}_{\rm cr}$ of the critical
solution.  Moreover, for each value of ${{a}}$ in this range there are
two solutions, with the one having the smaller value of $(1/A)_{\rm
min}$ being unstable \cite{stability,stability2}.  Hence, for this
range of $b$ our methods cannot find the critical limit of the regular
monopole solutions.

\subsection{Approach to criticality: low--${{b}}$ case}

We investigate the approach to criticality by studying the nature of
the monopole solutions as ${{a}} = 8\pi Gv^2$ is increased with the
Higgs self-coupling held fixed.  Two distinct behaviors are seen,
depending on the size of ${{b}} = \lambda/e^2$.
\begin{figure} \begin{center}\PSbox{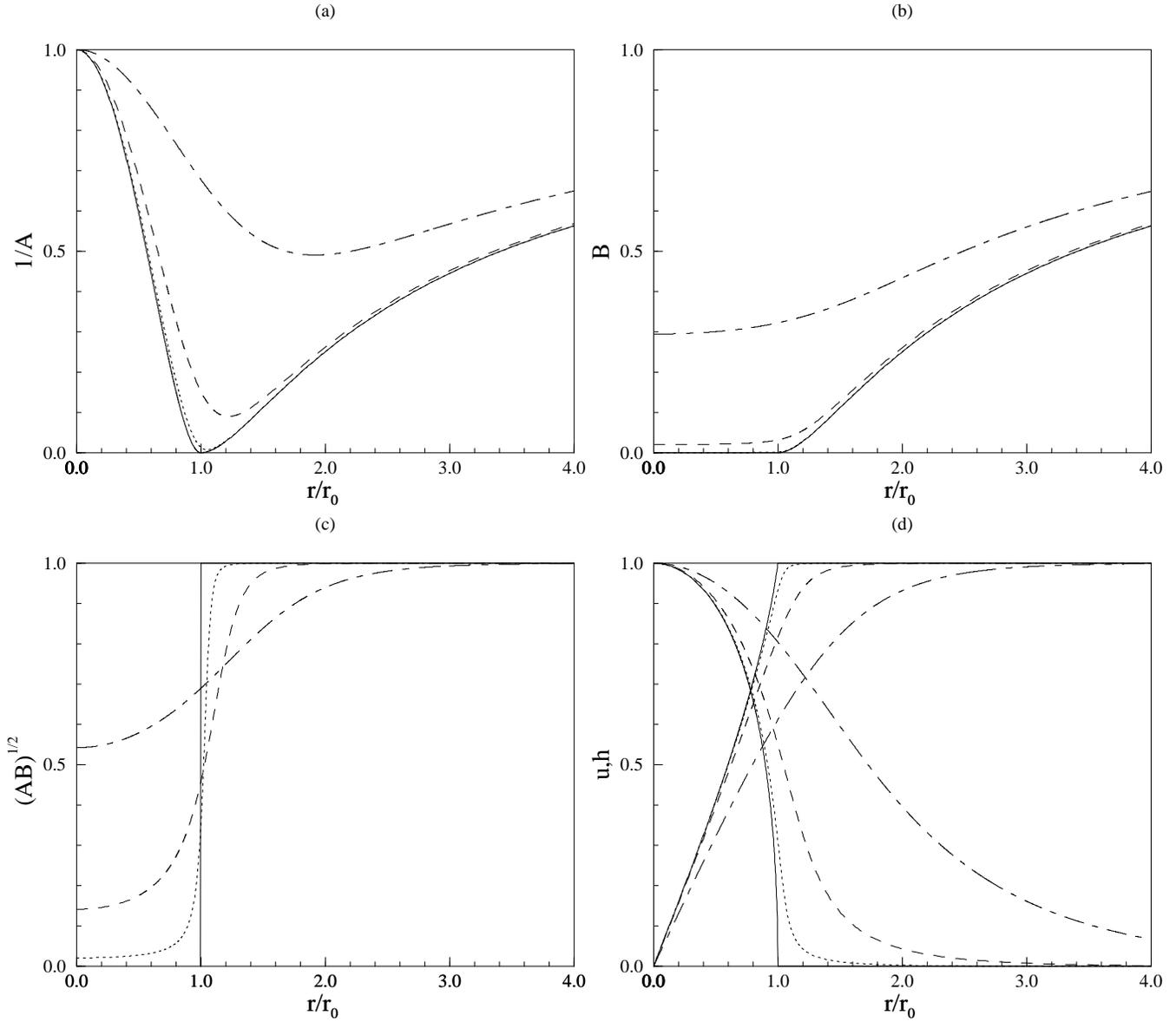
hscale=100 vscale=100 hoffset=-120 voffset=-35}{5in}{6in}\end{center}
\caption{
Monopole solutions for ${{b}} = 1.0$ and various values of $a$.  
The progression from
dot-dashed line, to dashed line, to dotted line, to solid line
corresponds to ${{a}} = 1.0, 2.0, 2.3,$ and $2.37$.  The first three
panels depict the metric functions (a) $1/A(r)$, (b) $B(r)$, (c)
$(AB)^{1/2}(r)$.  (d) Matter fields variables;  $u(r)$ begins at unity
at the origin and asymptotes to zero 
as $r \rightarrow \infty$.  $h(r)$ is zero at the origin and asymptotes
to unity as $r\rightarrow \infty$.} 
\label{fig:low}
\end{figure}
We will describe the low-$b$ behavior in detail in this subsection,
and the high-$b$ behavior in the next.  Figure~\ref{fig:low}
illustrates the behavior for ${{b}}=1.0$, a typical low-$b$ case.  As
${{a}}$ increases and gravitational effects become stronger, $1/A(r)$
begins to dip down until at ${{a}} = {{a}}_{\rm cr}$ it develops a
double zero, corresponding to an extremal horizon, at $r=r_0$, the
horizon radius of the extremal Reissner-Nordstrom solution.  At the
same time, the matter fields $u(r)$ and $h(r)$ are pulled inward.  At
${{a}} = {{a}}_{\rm cr}$, the variation of these fields occurs
entirely within the horizon.  Only the Abelian Coulomb magnetic field
survives outside the horizon, while the metric for $r\ge r_0$ is
precisely that of the extremal Reissner-Nordstrom black hole.

All of this reproduces results found in earlier work
\cite{orig1,ortiz,orig2,orig2b}.  Some features that were not
previously stressed are revealed when we examine $B(r)$.  Coming in
from large $r$, $B$ decreases with $1/A$ until the latter reaches its
minimum.  $B$ then continues to decrease, although at a much smaller
rate.  For very small ${{b}}$ this decrease continues all the way in
to $r=0$, while for somewhat larger values of ${{b}}$ there is a
minimum in $B$ at a finite $r < r_0$.  The situation for the critical
solution is somewhat ambiguous.  If we adopt the conventional
normalization $B(\infty)=1$, then $B(r)$ vanishes identically inside
the horizon.  If instead we set $B(0)=1$, then $B$ is finite and
varying inside the horizon and infinite outside the horizon, with a
minimum either at $r=0$ or at some finite radius, depending on the
value of ${{b}}$.  In neither case is the minimum a zero of $B$.

\begin{figure} \begin{center}\PSbox{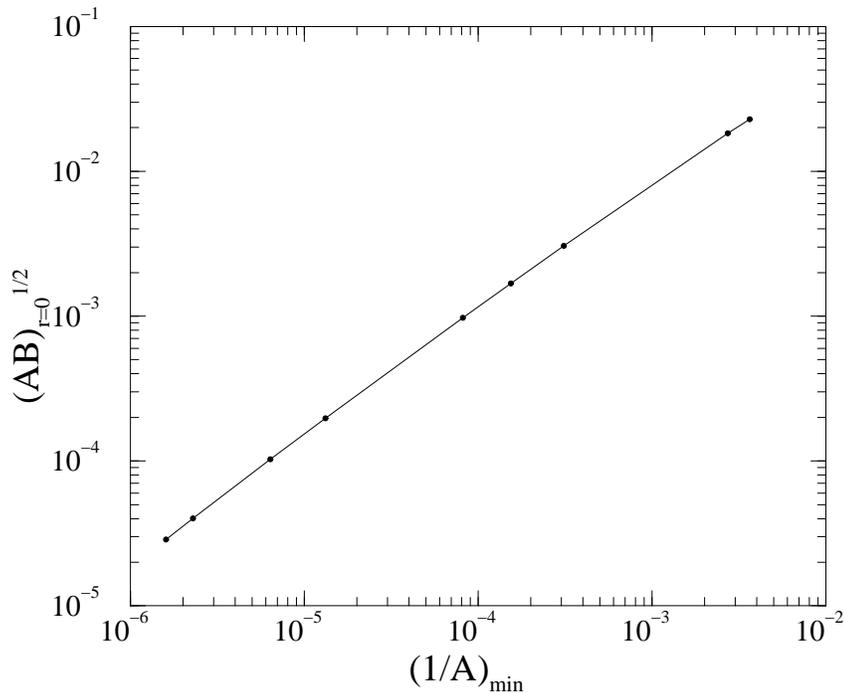
hscale=65 vscale=65 hoffset=-25 voffset=-35}{5in}{3in}\end{center}
\caption{$(AB)_{r=0}^{1/2}$ versus $(1/A)_{\rm min}$ for ${{b}} = 16.0$
and various values of ${{a}}$ near ${{a}}_{\rm cr}$.
}
\label{fig:AB}
\end{figure}

Closely related to this is the behavior of $\sqrt{AB}$.  When ${{a}}$ is small,
this is very nearly  constant, with a value close to unity.  As ${{a}}$ is
increased,  $\sqrt{AB}$ develops a step-like behavior until, at criticality, it
is precisely a step function centered at the horizon.   This is in sharp
contrast with the Schwarzschild and Reissner-Nordstrom solutions,  where
$\sqrt{AB} =1$ everywhere.  To show how this behavior becomes more
pronounced as the solution approaches criticality, in  Figure~\ref{fig:AB} we
plot the value of $\sqrt{AB}$ at the origin as a function of $(1/A)_{\rm min}$
for ${{b}}=16$.  Note that there is a power law relationship between the two. 
Similar curves, but with different powers, are found throughout the low-${{b}}$
regime.

\begin{figure} \begin{center}\PSbox{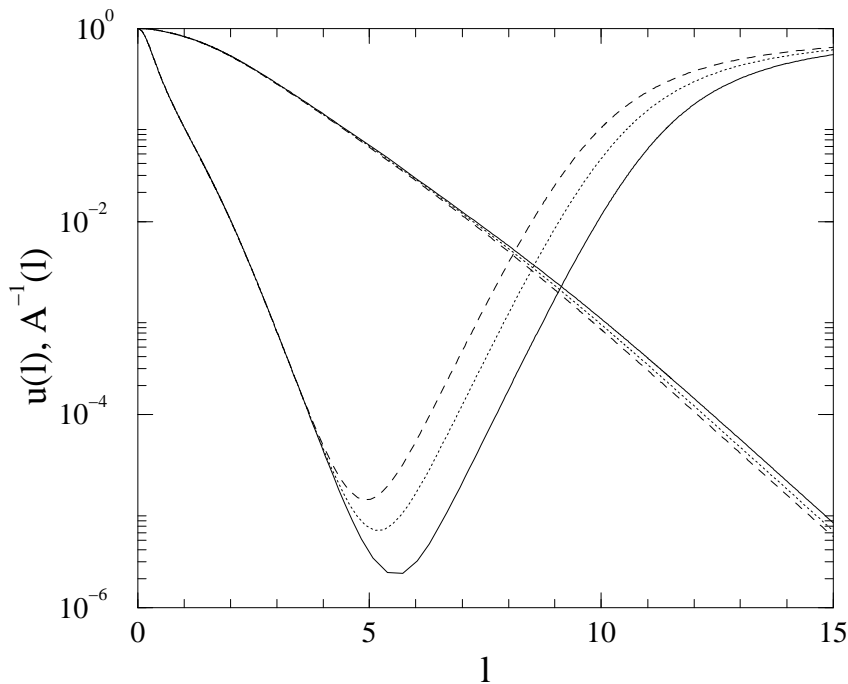
hscale=65 vscale=65 hoffset=-25 voffset=-35}{5in}{3in}\end{center}
\caption{$A^{-1}$ (curves with minima) and 
$u$ (monotonic curves) as functions of the
proper length $l$ for ${{b}} = 16.0$.  The progression from dashed
line, to dotted line, to solid line corresponds to ${{a}} = 1.529$,
$1.53$ and $1.531$ with ${{a}}_{\rm cr} \approx 1.532$.  }
\label{fig:near}
\end{figure}

As a final illustration of the behavior of the low-${{b}}$ solutions near
criticality, in Fig.~\ref{fig:near} we plot $u$ for several
near-critical solutions.   However, rather than using the variable $r$, we have
plotted $u$ as a function of the proper length
\begin{equation}
	l(r) = \int_0^r dr\ \sqrt{A(r)} \, .
\label{properdist}
\end{equation}
We see that, for sufficiently large $l$, $u(l)$ is close to a decaying
exponential, and that this behavior does not significantly change as one
crosses the minimum of $1/A$; similar behavior is seen with $1-h$.

\subsection{Approach to criticality: high--${{b}}$ case}

\begin{figure} \begin{center}\PSbox{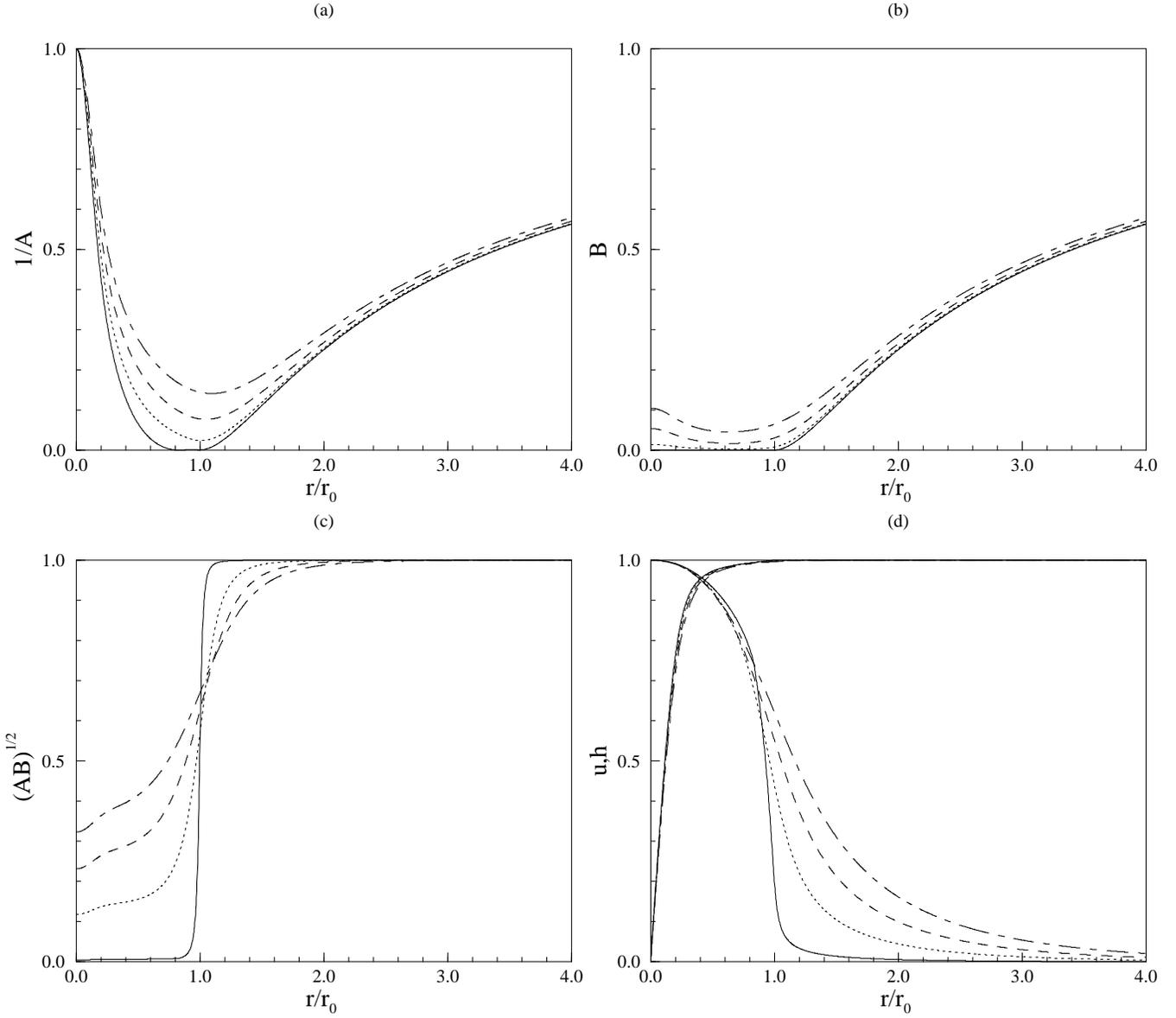
hscale=100 vscale=100 hoffset=-120 voffset=-35}{5in}{6in}\end{center}
\caption{ Monopole solutions for ${{b}} = 100$ and various values of $a$.  
The progression from
dot-dashed line, to dashed line, to dotted line, to solid line
corresponds to ${{a}} = 1.0$, 1.1, 1.2,
and $1.29344$.  The first three 
panels depict the metric functions (a) $1/A(r)$, (b) $B(r)$, (c)
$(AB)^{1/2}(r)$.  (d) Matter fields variables;  $u(r)$ begins at unity
at the origin and asymptotes to zero 
as $r \rightarrow \infty$.  $h(r)$ is zero at the origin and asymptotes
to unity as $r\rightarrow \infty$.}
\label{fig:high}
\end{figure}

The scenario when ${{b}}$ is large differs significantly from that for
small ${{b}}$; we illustrate this in Fig.~\ref{fig:high} for the case
of ${{b}}=100$. So long as ${{a}}$ is not too near its critical value,
the qualitative evolution of the metric functions and field variables
is similar to that in the previous case: $1/A(r)$ dips down in a
region $r \approx r_0$, the value of $B(r)$ decreases in a region near
the origin, and the fields $u, h$ are drawn into the core of the
system.

\begin{figure} \begin{center}\PSbox{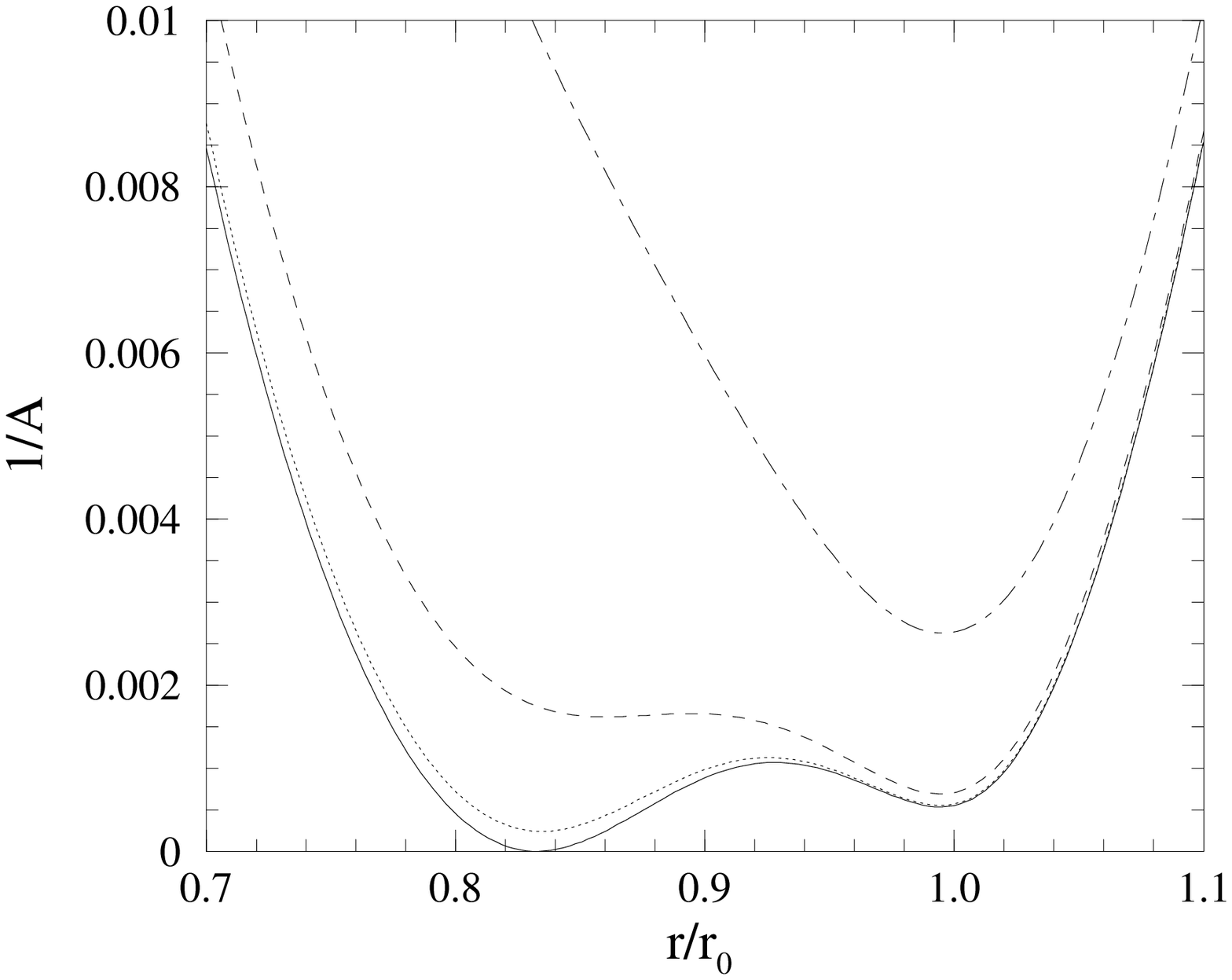
hscale=65 vscale=65 hoffset=-25 voffset=-35}{5in}{3in}\end{center}
\caption{Details of the metric function $1/A(r)$ near criticality at
${{b}} = 100$.  The progression from dot-dashed line, to
dashed line, to dotted line, to solid line corresponds to ${{a}} =
1.27$, 1.29, 1.293, and $1.29344$.}
\label{fig:dc}
\end{figure}

\begin{figure} \begin{center}\PSbox{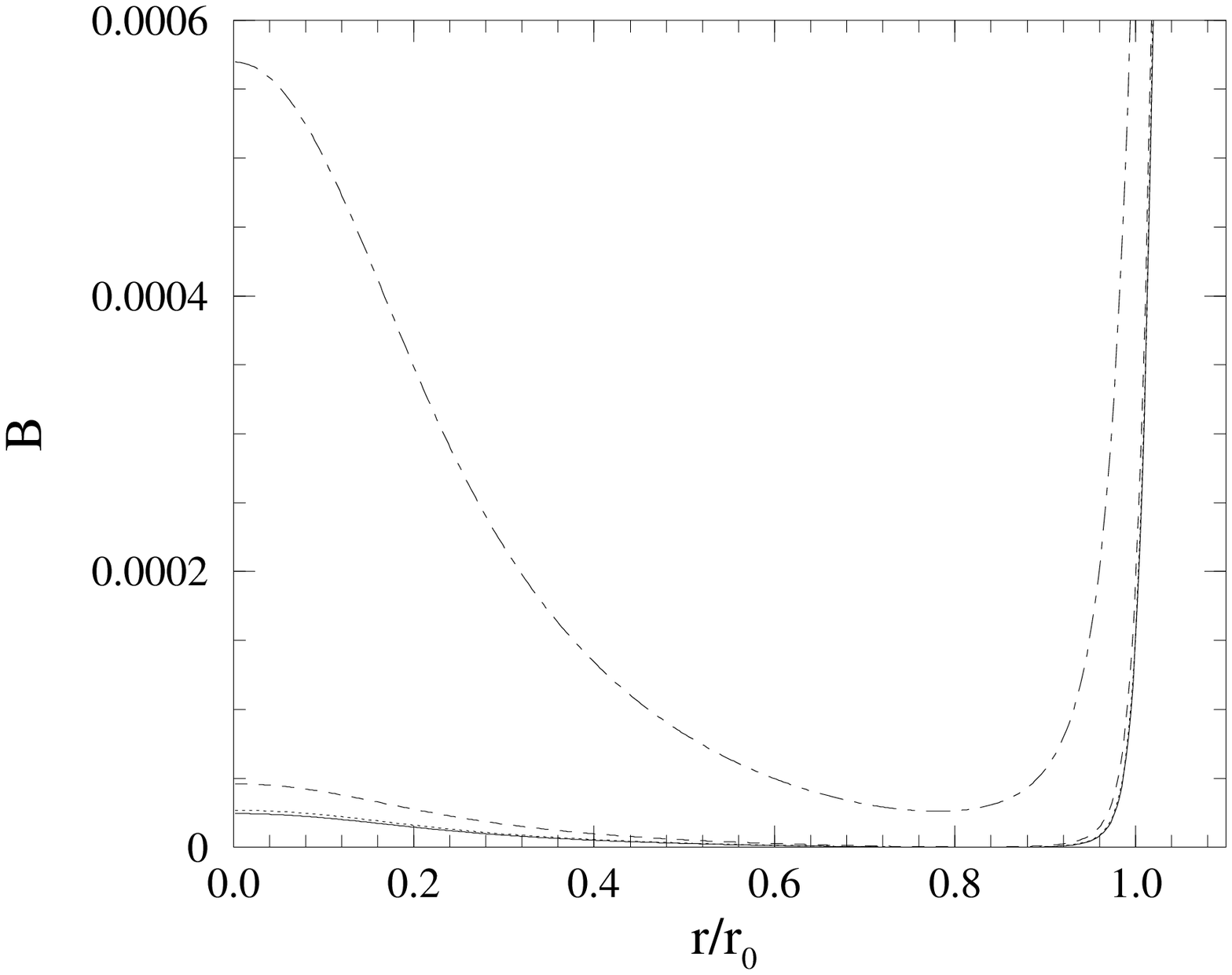
hscale=65 vscale=65 hoffset=-25 voffset=-35}{5in}{3in}\end{center}
\caption{Details of the metric function $B(r)$ near criticality at
${{b}} = 100$.  The progression for different ${{a}}$'s is the same
as in Figure 5.}
\label{fig:db}
\end{figure}

Near criticality, a different behavior emerges. The evolution of
$1/A(r)$ as the solution approaches criticality is depicted in
Fig.~\ref{fig:high}a and, in more detail, in Fig.~\ref{fig:dc}.
Although initially similar to that the for the low-${{b}}$ case, 
the decrease of the minimum at $r\approx r_0$  ceases before
it actually reaches zero.  As this occurs, a second minimum at a
radius $r<r_0$ rapidly drops down and forms a double zero,
corresponding to an extremal horizon, at $r = r_* < r_0$.
Figure~\ref{fig:high}b (and, in more detail, Fig.~\ref{fig:db})
shows the function $B(r)$ as ${{a}}$ approaches
its critical value.  In contrast to the small-${{b}}$ case, in the
critical limit the 
minimum of $B$ is a zero located at 
the horizon, $r=r_*$.  


The evolution of $\sqrt{AB}$ is shown in 
Fig.~\ref{fig:high}c.  The behavior is similar to that for the 
small-${{b}}$ case until the outer minimum of $1/A$
stops decreasing and the inner minimum begins to appear.  
Until this point, there is a power-law relationship between
$\sqrt{AB}|_{r=0}$
and $(1/A)_{\rm min}$ similar to that for the small-${{b}}$ case, with
most of the variation in $\sqrt{AB}$ occurring at $r\approx r_0$.
Once the inner minimum begins to appear,
the decrease in $\sqrt{AB}|_{r=0}$ slows down, so that even for
the critical solution $\sqrt{AB}|_{r=0}\ne 0$.   

Figure~\ref{fig:high}d depicts the matter fields $u(r)$ and $h(r)$ for
a series of values of ${{a}}$ at fixed large ${{b}}$.  Their behavior
is analogous to that for small ${{b}}$ except that the fields are
never completely drawn into the region $r<r_0$; the degree to which
they are drawn into this region is dictated by the value of $1/A$ at
the outer minimum at criticality.  The smaller that minimum value of
$1/A$, the more contained the matter fields are.  Since they have
nontrivial fields outside the horizon the critical solutions for large
${{b}}$ are examples of extremal black holes with non-Abelian hair.

\begin{figure} \begin{center}\PSbox{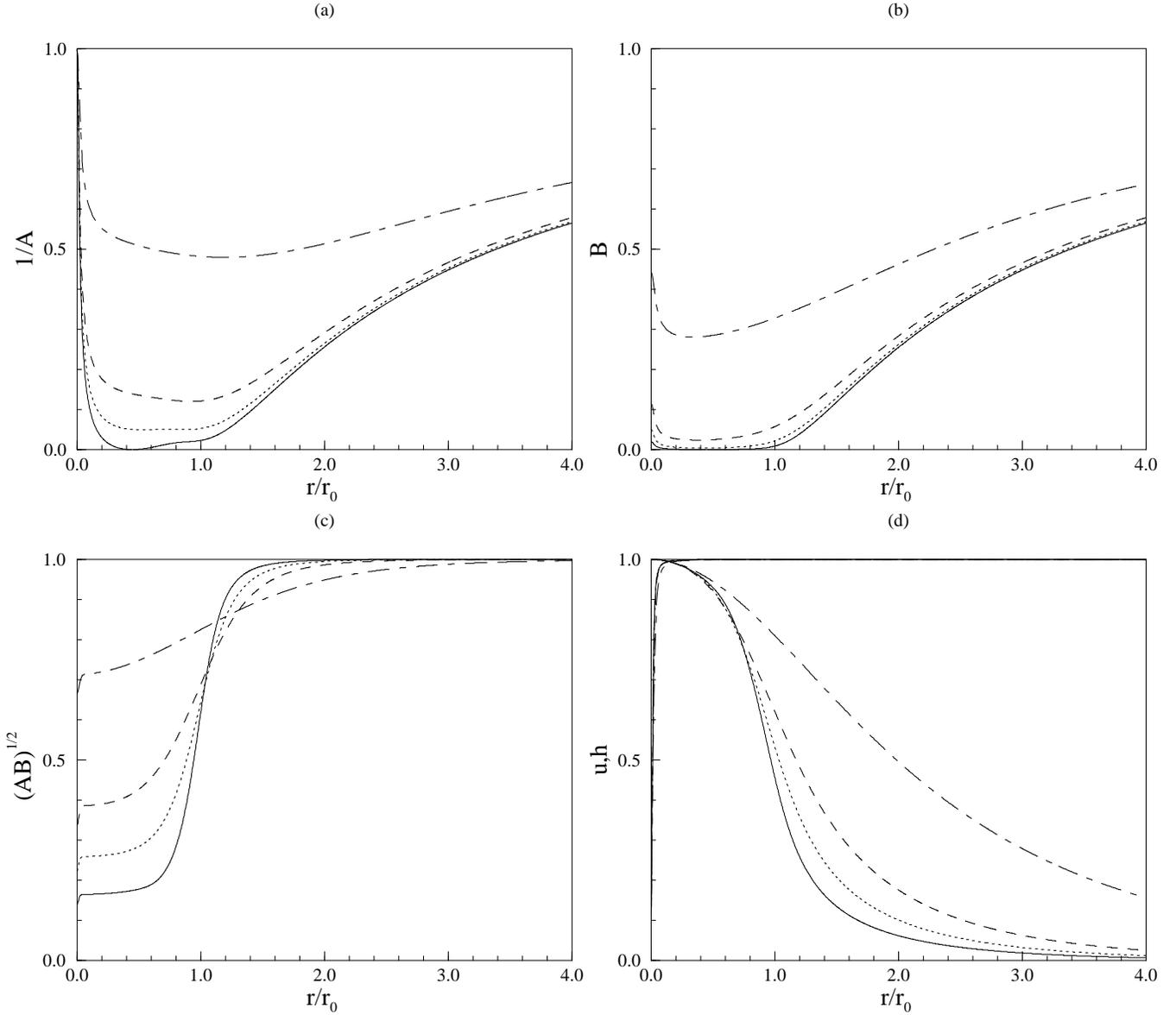
hscale=100 vscale=100 hoffset=-120 voffset=-35}{5in}{6in}\end{center}
\caption{ Monopole solutions for ${{b}} = 1000$ and various values of $a$.  
The progression from
dot-dashed line, to dashed line, to dotted line, to solid line
corresponds to ${{a}} =  0.5$, 0.9, 1.0,
and $1.05576$.  The first three 
panels depict the metric functions (a) $1/A(r)$, (b) $B(r)$, (c)
$(AB)^{1/2}(r)$.  (d) Matter fields variables;  $u(r)$ begins at unity
at the origin and asymptotes to zero 
as $r \rightarrow \infty$.  $h(r)$ is zero at the origin and asymptotes
to unity as $r\rightarrow \infty$.}
\label{fig:big}
\end{figure}
The qualitative picture differs somewhat when ${{b}} \simge 400$, in
that the $1/A$ always has only a single minimum.  One may think of the
sequence of monopole solutions here as large--${{b}}$ solutions in
which the inner minimum of $1/A$ drops out sufficiently early that no
double minimum solutions exist.  To illustrate this, in
Fig.~\ref{fig:big} we show the approach to criticality for
${{b}}=1000$.

\subsection{Behavior of the critical solutions}

The critical solutions themselves are not accessible through our
numerical method.  Nevertheless, we can obtain good approximations to
these by following sequences of regular monopole solutions.  Here we
briefly summarize how some of the properties inferred from these
numerical solutions vary with $b$. 

\begin{figure} \begin{center}\PSbox{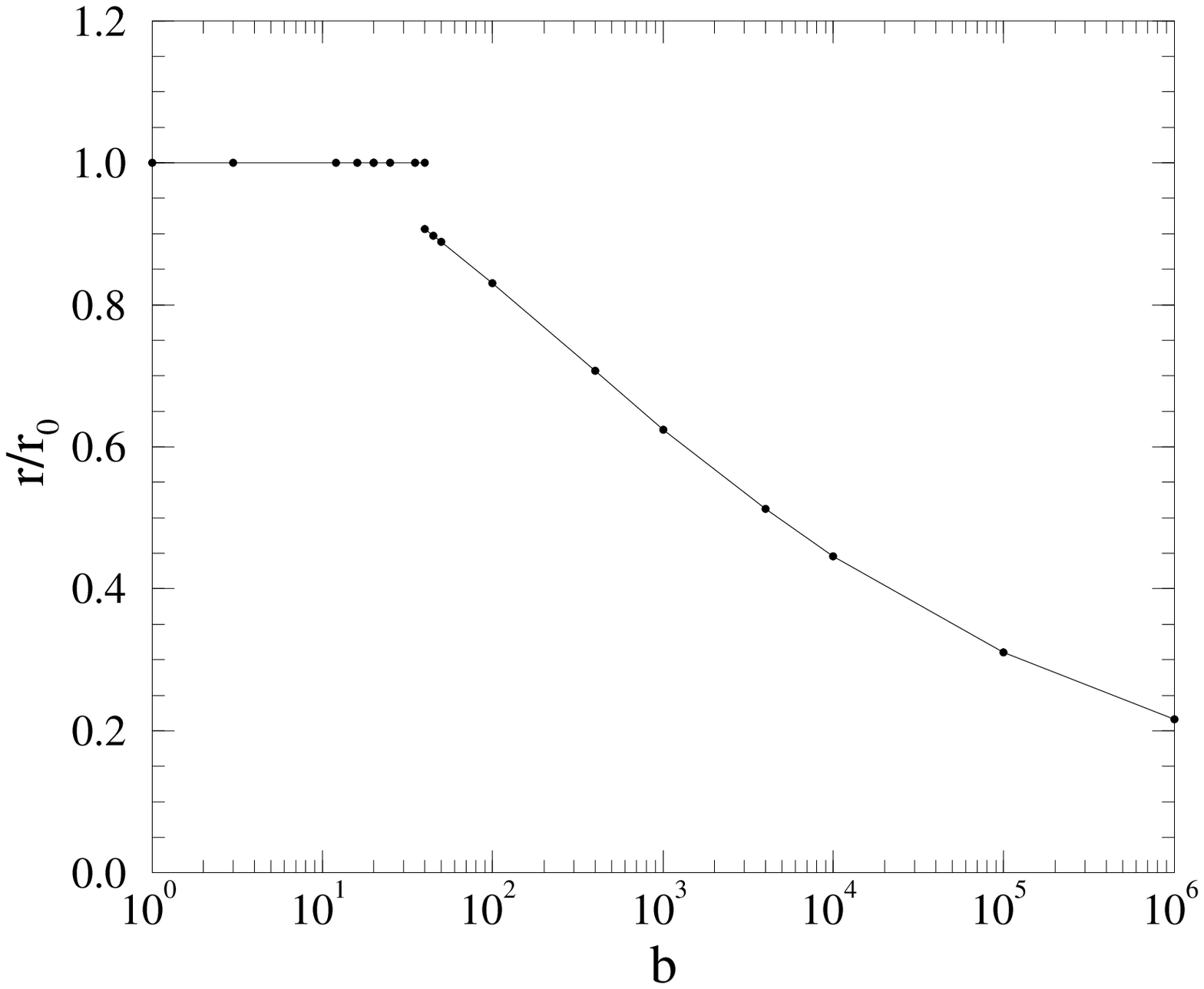
hscale=65 vscale=65 hoffset=-25 voffset=-35}{5in}{3in}\end{center}
\caption{Extremal horizon radius for various values of ${{b}}$.
Analytic arguments indicate that $r_* \rightarrow 0$ as $b
\rightarrow \infty$.} 
\label{fig:radius}
\end{figure}
As we have already noted, the critical solutions for small $b$ have
horizons at the Reissner-Nordstrom radius $r_0$, while for large $b$
the horizon occurs at a value $r_* < r_0$ (see Fig.~\ref{fig:radius}).
There is a discontinuity in passing from the low-$b$ to the high-$b$
regime: The critical solution does not continuously vary from one type
to the other, but instead undergoes something like a first-order
transition.  For a value of $b$ just below the transition, an inner
minimum appears\footnote{This inner minimum in low-$b$ monopole
solutions only appears for $b$ very close to the transition point and
when $a$ is very near $a_{\rm cr}$.} at a radius $r < r_0$, while
$1/A$ has a double zero at $r_0$.  As $b$ increases, the inner minimum
rapidly descends and at the transition $1/A$ has two degenerate
minima, one at $r_* < r_0$ and the other at $r_0$.  As $b$ increases
further the outer minimum moves upward, while the inner minimum at
$r_*$ remains at zero.  Eventually the outer minimum vanishes, and the
only minimum in $1/A$ is associated with the horizon at $r_*$.  This
radius $r_*$ asymptotes to zero as $b \rightarrow \infty$.  As $b$ is
increased, the critical black hole solution approaches that found in
the literature in the infinite $b$ limit \cite{orig2,AB}.

\begin{figure} \begin{center}\PSbox{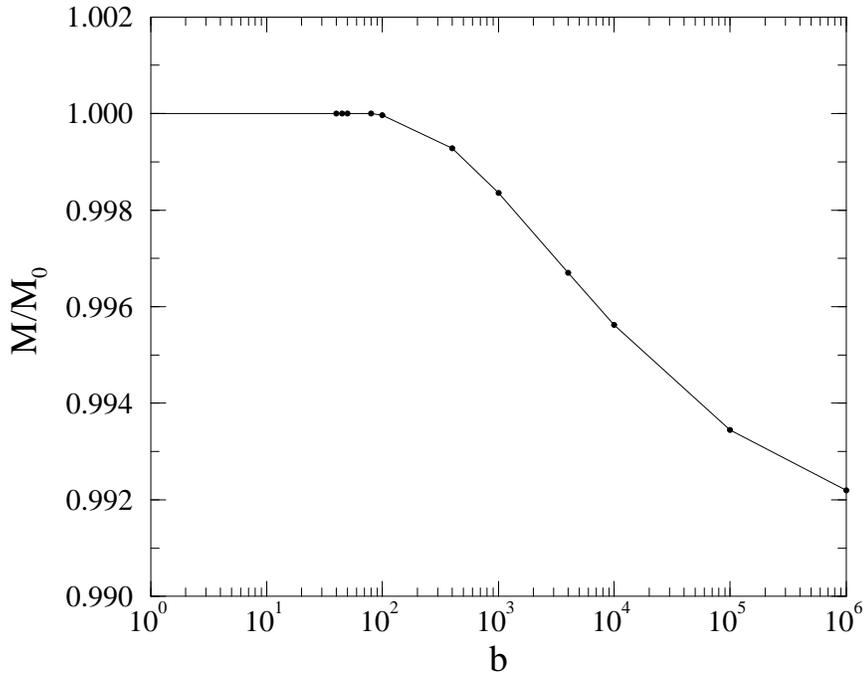
hscale=65 vscale=65 hoffset=-25 voffset=-35}{5in}{3in}\end{center}
\caption{Mass of critical solutions as a function of ${{b}}$.
Low-${{b}}$ type solutions (${{b}} < 40$) all have $M=M_0$.  The dots
indicate the values for high-$b$ type solutions.}
\label{fig:mass}
\end{figure}
\pagebreak

The mass of the critical solution can be inferred from the long-range
behavior of $1/A$.  In Fig.~\ref{fig:mass} we plot this mass as a
function of ${{b}}$.  As expected, the mass of the low-${{b}}$
solutions is just that of an extremal Reissner-Nordstrom black hole.
However, in the high-${{b}}$ regime the mass of the critical solution
is less than the extremal Reissner-Nordstrom value $M_0$.  The mass
decreases with increasing ${{b}}$, and appears to have an asymptotic
value of about 0.990.  Note that there is no discontinuity in the mass
in going from the low-$b$ to the high-$b$ regime.

\begin{figure} \begin{center}\PSbox{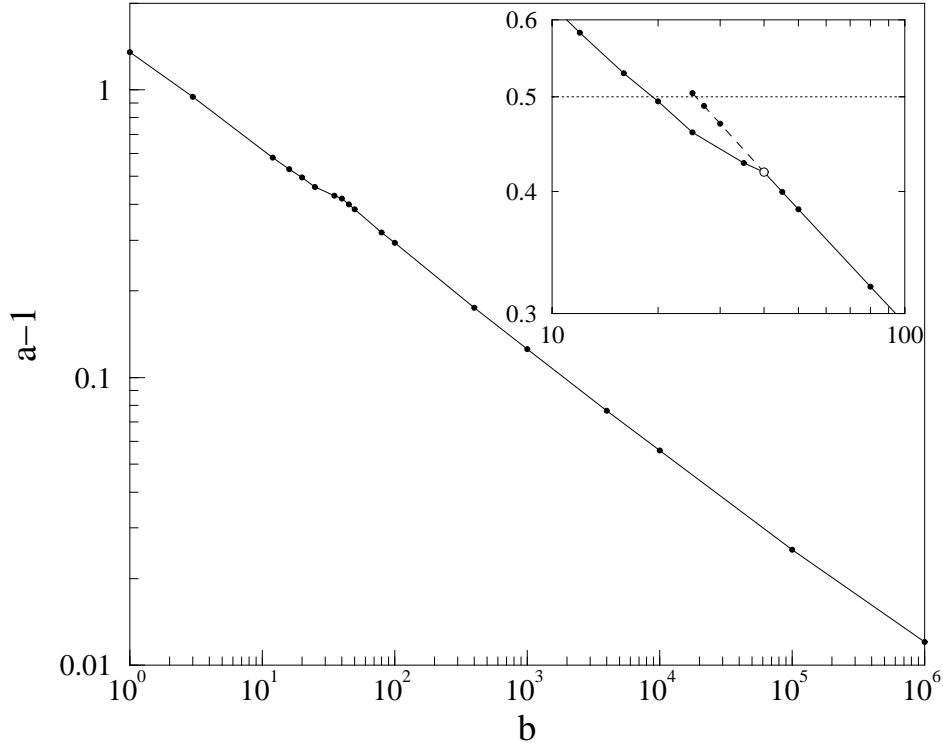
hscale=65 vscale=65 hoffset=-25 voffset=-35}{5in}{3.5in}\end{center}
\caption{Critical values for ${{a}}-1$ as function of ${{b}}$.  The
inset shows in detail the transition region between low-${{b}}$ and
high-${{b}}$ behavior, with the open circle indicating the apparent
transition point at ${{b}} = 40$.  Note that at this point
${{a}}_{\rm} = 1.4187 < 1.5$, the latter value indicated by the dotted
line.  The dashed line is an extension of the high-$b$ type critical
solutions to smaller $b$.  For the solutions represented by this
dashed line, the fields are well-behaved inside the horizon, but are
not asymptotically flat at spatial infinity.}
\label{fig:crit}
\end{figure}
Finally, Fig.~\ref{fig:crit} shows ${{a}}_{\rm cr}$ as a function of
${{b}}$.  The critical value ${{a}}_{\rm cr}$ is always of order unity
and is a monotonically decreasing function of ${{b}}$.  This can be
understood by recalling that the mass of the flat space monopole
increases (by a factor of 1.8) as the ratio of the Higgs mass to the
vector mass varies varies from zero to infinity.  Stated differently,
the value of $v$ needed to achieve a given mass decreases with
increasing Higgs mass, making it plausible that the critical $v$
should decrease in a similar fashion.

For large ${{b}}$ there appears to be a power-law relationship between
${{b}}$ and ${{a}}_{\rm cr}-1$.  That ${{a}}_{\rm cr} \rightarrow 1$
as $b \rightarrow \infty$ is consistent with observations made in past
investigations \cite{orig2,orig2b,AB}.  The literature also
\cite{orig2} indicates that the curve should asymptote to ${{a}}_{\rm
cr} \approx 3.94$ as ${{b}}\rightarrow 0$.  The kink in the data at $b
\approx 40$ is real and reflects the transition between the small and
large ${{b}}$ regimes.  Note that $a_{\rm cr} < 1.5$ at this
transition point; this fact will be of importance later.  The dashed
curve in the inset of Fig.~\ref{fig:crit} shows an extension of the
high-$b$ type critical solutions into the low-$b$ regime.  The
corresponding solutions are well behaved inside the horizon, but are
not asymptotically flat at spatial infinity; see the Appendix for more
details.

\section{Analytic constraints from extremal horizons}

We would now like to use analytic methods to gain some deeper
understanding of our numerical results.  We begin by focusing on the
critical solutions with extremal horizons, but regular at the origin,
that are the limits of the gravitating monopole solutions.  In this
section we will obtain a set of conditions at the extremal horizon
that are necessary, although not sufficient, for the existence of such
solutions.  These turn out to allow two distinct types of limiting
solutions.  We discuss both of these, as well as the nearby nearby
nonsingular solutions, in more detail in the next section.  Our
analysis of the behavior of the fields near the horizon complements
and extends earlier treatments \cite{orig2b,hajicek}.

Because a simultaneous zero of both $1/A$ and its first derivative is
a singular point of the differential Eqs.~(\ref{ueqn}) and
(\ref{heqn}), the critical solutions are nonanalytic at the horizon radius
$r_*$.  Ordinarily, physical considerations would constrain the
allowable singularities at such a point.  However, we are not actually
requiring that the solution be physically acceptable, but only that it
be the limiting point of a family of physically acceptable solutions.
Keeping this in mind, it seems reasonable to impose the following set
of requirements:

\begin{enumerate}
	\item We assume that the functions $u$, $h$, $1/A$, and $B$ are all
finite and continuous at $r_*$.
	\item Since $r$ is a singular coordinate at the horizon, we should not
assume that the derivatives of these functions with respect to $r$ are
continuous, or even finite, at $r_*$.  Instead, we require only that
$d(1/A)/dr$ vanishes at the horizon and that $du/dr$ and $dh/dr$
diverge less rapidly than $A^{1/2}$ does.
	\item We assume that the leading singularities of these quantities near
the horizon can be approximated by (not necessarily integer) powers of
$|r-r_*|$, although we allow for the possibility that both the power
and the coefficient of the leading term may be different on opposite
sides of the horizon.
\end{enumerate}

Given these assumptions, Eqs.~(\ref{ueqn}) and (\ref{heqn}) imply that the
matter fields at the horizon, which we denote by $u_*$ and $h_*$, must lie at
one of the stationary points of $U$ that were enumerated in Sec.~II.  Once
these are specified, $r_*$ is determined by Eq.~(\ref{Meqn}), which reduces to
\begin{equation}
  0 = 1 - 8\pi Gr_*^2U(r_*)\ .
\label{Uathorizon}
\end{equation}

The monopole solutions without horizons were solutions to a boundary
value problem with conditions imposed at $r=0$ and $r=\infty$.  The
existence of the extremal horizon imposes three more conditions (on
the values of $u$, $h$, and $1/A$ at $r=r_*$) and thus leads to a
pair of boundary value problems, one for the interval $0 \le r \le
r_*$ and one for $r_* \le r <\infty$, that must be solved
simultaneously.  The interior problem has three boundary conditions at
the origin and three at the horizon, for a total of six, while the
exterior has two at spatial infinity and three at the horizon, giving
five in all.

A standard approach to such problems is to look for a family of
solutions obeying the conditions at one of the boundaries.  If these
can be shown to depend on $N$ adjustable parameters, integration of
these solutions will (assuming no singularities intervene) give an
$N$-parameter family of solutions in the neighborhood of the other
boundary.  Generically, a necessary, although not sufficient,
condition for a solution is that $N$ be greater than or equal to the
number of conditions imposed at the second boundary.  Thus, at first
thought it would seem that, given a suitable set of values for $u_*$,
$h_*$, and $r_*$, we would need a two-parameter family of
solutions as $r- r_* \rightarrow 0_+$ to solve the
exterior boundary value problem and a three-parameter family for $r-
r_* \rightarrow 0_-$ to solve the interior problem.  However, we do
not expect to find extremal solutions for arbitrary values of
parameters.  Instead, for fixed ${{b}}$ we must adjust $v$ (or,
equivalently, ${{a}}$) to its critical value.  We may therefore view
$v$ as an extra adjustable parameter, and need only require the
existence of two-parameter families of solutions on both sides of the
horizon.  The nonanalyticity at the horizon allows us to choose the
parameters on the two sides independently.

There are several caveats.  First, the presence of the
appropriate number of adjustable parameters at the horizon does not
guarantee the existence of a solution.  Global considerations that are
beyond the scope of this analysis may make it impossible to satisfy
all of the boundary conditions.  Second, it may, and in some cases
does, happen that the leading behavior near the horizon is fixed and
that 
the adjustable parameters appear only in subdominant terms.
Finally, if the values of $u$ and $h$ at the horizon are the same
as at the origin (infinity), the interior (exterior) boundary value
problem for the matter fields has a trivial solution, and so it is not
necessary to have any adjustable parameters.

We find it convenient to define
\begin{equation}
   \psi = \left(\matrix{ u-u_* \cr \cr er_*v(h-h_*)/\sqrt{2} } \right)
\end{equation}  
and to use the dimensionless position variable $x \equiv (r-r_*)/r_*$.   
It is also useful to define the matrix
\begin{equation}
     {\cal M}_{ij} ={e^2 r_*^4 \over 2}  \left.{\partial^2 U\over
             \partial\psi_i 
             \partial\psi_j} \right|_{r_*}\ .
\end{equation}
The orthonormal eigenvectors and the eigenvalues of ${\cal M}$ play an
important role in the analysis; we denote these by $\theta_a$ and
$\mu_a$ ($a=1$, 2), respectively.  Finally, we define the
two-component vector
\begin{equation} 
     \Phi_i = {e^2 r_*^5 \over 2} \left.{\partial^2 U\over \partial r
            \partial\psi_i} \right|_{r_*} 
\end{equation}
as well as the ratio
\begin{equation}
     \sigma \equiv {4\pi G \over e^2 r_*^2} = {r_0^2\over r_*^2}
\end{equation}
where, as before, $r_0$ is the horizon radius of the extremal
Reissner-Nordstrom black hole.

Equation~(\ref{Meqn}) and the equations obtained by substitution of
Eq.~(\ref{ABeqn}) into Eqs.~(\ref{ueqn}) and (\ref{heqn}) form a set of three
equations for the functions $u$, $h$, and $1/A$.  In the notation we have just
defined, the first two of these can be compactly written as
\begin{equation} 
    {1\over A} \psi'' + \left({1\over A}\right)' \psi'
     + {2\over A} \sigma  [(\psi')^t \psi'] \psi'
      =  {\cal M} \psi  +  \Phi x + \cdots
\label{psieqn}
\end{equation}
while the third becomes 
\begin{equation}
     \left({1\over A}\right)' 
        = 2Fx - 2\sigma
     \left[  \psi^t{\cal M} \psi
                   + {1\over A} (\psi')^t \psi' \right] + \cdots
\label{newMeqn}
\end{equation}
where  
\begin{equation}
     F = -\sigma  e^2 r_*^3 \left.{\partial (r^2 U) \over \partial
           r}\right|_{r_*}\ .
\end{equation}
Here primes denote derivatives with respect to $x$, while the ellipses
represent terms, of higher order in either $x$ or the components of
$\psi$, that can be neglected here.
 
We are assuming that the leading behavior of the various functions can
be approximated by powers of $x$. It is then fairly easy to
show\footnote{Assume that $1/A \sim |x|^\alpha$.  If $\alpha>2$,
Eq.~(\ref{newMeqn}) is dominated by the two terms not involving $1/A$,
and one finds that $\psi \sim |x|^{1/2}$.  This in turn implies that
Eq.~(\ref{psieqn}) is dominated by a single term, the first one on the
right hand side, and hence has no solution.  If instead $\alpha<2$,
the two terms involving $1/A$ dominate Eq.~(\ref{newMeqn}), again
implying that $\psi \sim |x|^{1/2}$.  Equation~(\ref{psieqn}) is now
dominated by the three terms on the left hand side.  Because the last
two of these cancel, there is again no solution.} that  
\begin{equation} 
     {1\over A} = k x^2 + \cdots
\end{equation}
This reduces Eqs.(\ref{psieqn}) and (\ref{newMeqn}) to 
\begin{equation} 
   k\left[ x^2 \psi'' + 2x \psi' + 
         2\sigma x^2 [(\psi')^t \psi']
        \psi' \right] 
      =  {\cal M} \psi  +  \Phi x + \cdots
\label{koneeqn}
\end{equation}
and
\begin{equation}
      k = F - \sigma
     \left[ {1 \over x }\psi^t{\cal M} \psi + kx (\psi')^t \psi'  
           \right] + \cdots 
\label{ktwoeqn}
\end{equation}

We now turn to the behavior of $\psi$ as $x\rightarrow 0$.  We see that
$\psi$ must vanish at least as fast as $|x|^{1/2}$, 
since otherwise the nonlinear term on the left-hand side of the
equation dominates and there is no solution.  
Hence, we may write
\begin{equation}
    \psi = \eta_{1/2} |x|^{1/2}  + \left[ \eta_1 x + \eta_2 x^2 +
\cdots \right] + \eta_{\gamma_1} |x|^{\gamma_1}
            + \eta_{\gamma_2} |x|^{\gamma_2} + \cdots
\label{psiexp}
\end{equation} 
where the ellipsis within the square brackets represents higher-order
analytic terms, the $\gamma_j >1/2$ are noninteger powers to 
determined, and the final ellipsis represents smaller nonanalytic terms
that are determined by the lower-order terms.

When this expansion is substituted into Eq.~(\ref{koneeqn}), the terms of order
$|x|^{1/2}$ give the nonlinear equation 
\begin{equation}
   k \left[{3\over 4} -\epsilon {\sigma \over 4}  
      \left(\eta_{1/2}^t \eta_{1/2}\right) \right]\eta_{1/2} 
     =  {\cal M} \eta_{1/2}
\end{equation} 
where $\epsilon$ is equal to plus or minus unity according to whether
$x$ is negative or positive (i.e., for the interior and exterior
problems, respectively).  A solution is possible only if $\eta_{1/2}$
is proportional to one of the eigenvectors of ${\cal M}$.  Let us
denote this eigenvector by $\theta_\parallel$ and the orthogonal
eigenvector by $\theta_\perp$, with analogous conventions for the
eigenvalues.  We thus have
\begin{equation}
     \eta_{1/2}  = p \theta_\parallel
\end{equation}
with $p$ obeying
\begin{equation}  
    kp \left[ {3\over 4} -\epsilon {\sigma \over 4} p^2 \right]
            =  \mu_\parallel p\ .
\label{peqn}
\end{equation}
In addition, substitution of our expansion for $\psi$ into Eq.~(\ref{koneeqn})
yields  
\begin{equation}
   k  \left[ 1 -\epsilon {\sigma \over 4}   p^2 \right]  
     =  F  +\epsilon \sigma  \mu_\parallel p^2\ .
\label{keqn}
\end{equation}

These equations can be used to find $k$ as a function of $p^2$, and
the result then substituted back into Eq.~(\ref{peqn}).  For the
interior solution ($\epsilon=1$), this leads to the result that 
\begin{equation} 
    p^2_{\rm int} =  {F\over 2 \sigma \mu_\parallel} \left[  { 4
      \mu_\parallel \over F} -1 
      \pm   \sqrt{ 1 + { 4 \mu_\parallel \over F} } \right]   \qquad\quad {\rm
      or} \qquad\quad p^2_{\rm int} = 0
\label{pformula}
\end{equation}
and 
\begin{equation}
    k_{\rm int} = {4 F \over(2 -\sigma p_{\rm int}^2) ^2 }\ .
\label{kformula}
\end{equation}	
Since $k$ must be positive, we require that $F > 0$.  The requirement
that $p$ be real then implies that nonzero solutions for $p$ exist
only if $\mu_\parallel \ge -F/4 $; for the solution with the lower
choice of sign, there is the additional requirement that
$\mu_\parallel$ not lie between 0 and $3F/4$.

Although Eqs.~(\ref{peqn}) and (\ref{keqn}) allow a nontrivial
solution for $p^2$ in the exterior region, other constraints,
described below, require that
\begin{equation} 
    p^2_{\rm ext} = 0
\end{equation}
and hence that 
\begin{equation} 
    k_{\rm ext} = F\ .
\end{equation}

Having determined $\eta_{1/2}$, we can now turn to the remaining terms
in Eq.~(\ref{psiexp}).  Extracting the coefficients of the integral powers of
$x$ in Eq.~(\ref{koneeqn}) yields a series of inhomogeneous linear equations
that determine the $\eta_n$; the
first of these is\footnote{The
coefficient of $\eta_1$ on the left hand side of this equation
vanishes for certain choices of parameters, leaving $\eta_1$
undetermined.  A similar phenomenon can also happen in the equations
for the other $\eta_n$.  These parameter choices correspond to 
points where the power $\gamma_j$ of one of the nonanalytic terms goes
through an integer value.}
\begin{equation} 
  ( 2 k -  {\cal M}  ) \eta_1 
      -\epsilon k p^2 {\sigma \over 2}
   \left[ 2 \left(\theta_\parallel^t \eta_1\right)\theta_\parallel
          + \eta_1\right] 
    =  \Phi\ .
\label{linearcoeff}
\end{equation}

In a similar fashion, the nonanalytic $O(|x|^{\gamma_j})$ terms give a
linear equation for $\eta_{\gamma_j}$.  However, because the
inhomogeneous term in Eq.~(\ref{koneeqn}) is analytic and cannot
contribute, the resulting equation is homogeneous.  A solution of this
equation is possible only if $\eta_{\gamma_j}$ is proportional to one
of the $\theta_a$ and if $\gamma_j$ is a root of
\begin{eqnarray} 
 0&=& k \left[ \gamma(\gamma+1) 
      - {3\over 2}\epsilon\gamma  \sigma p^2  \right] -   \mu_\parallel 
     \, ,  \qquad  \eta_{1/2} \propto \theta_\parallel \cr 
0&=& k  \left[\gamma(\gamma+1) - {1\over 2}\epsilon\gamma  \sigma p^2  \right]
     -   \mu_\perp \, ,  \qquad 
              \eta_{1/2} \propto \theta_\perp\ .
\label{gammaeqn}
\end{eqnarray}
Note that both equations take the same form in the  
$p^2=0$ case, where the two $\mu_a$ are on the same footing.
Because the equation is homogeneous, $\eta_{\gamma_j}$ is determined
only up to an overall multiplicative constant.  In order that the
solution near the horizon have two adjustable constants, there must be
two independent nonanalytic terms with powers greater than 1/2.  Thus,
the pair of equations corresponding to the two $\mu_a$ must, between
them, have two roots greater than 1/2.

For $p=0$, the coefficient of $\gamma$ in Eq.~(\ref{gammaeqn}) is
positive, implying that at least one $\gamma$ must be negative for each
$\mu_a$.  In order to have two roots positive and greater than
1/2, we must require that both eigenvalues satisfy 
\begin{equation}
    \mu_a > {3 F\over 4} >0 \, , \qquad\qquad p=0\ .
\label{pzeromus}
\end{equation}

For $p \ne 0$, the form of Eq.~(\ref{gammaeqn}) depends on whether we
are considering the mode proportional to $\theta_\parallel$ or the one
proportional to $\theta_\perp$.  In either case, it is most convenient
to proceed by using Eq.~(\ref{peqn}) to eliminate $\mu_a$.  For
$\epsilon=-1$ the coefficient of $\gamma$ is positive, so at least one
root is negative for each mode.  Detailed examination of the
equation for the $\theta_\parallel$ mode then shows that it has no
roots greater than 1/2; since the equation for the $\theta_\perp$ mode
can have at most one such, we must set $p=0$ in the exterior region.
For $\epsilon=1$ (i.e., the interior region), the equation for the
$\theta_\parallel$ mode has one and only root greater than 1/2 for all
values of $\mu_\parallel$ consistent with the reality of $p$.  The
equation for the other mode can be rewritten as
\begin{equation}
    0 = \gamma\left(\gamma - {1\over 2}\right) 
       + \left(\gamma - {1\over 2}\right) \left( {2 \mu_\parallel
       \over k} \right) 
        + \left({\mu_\parallel -\mu_\perp\over k} \right)\ .
\end{equation}
For this to have a solution with $\gamma >1/2$, either $\mu_\parallel < \mu_\perp$
or $\mu_\parallel < 0$; since we will see that at most one of the $\mu_a$ can
be negative, the second alternative implies the first. Combining this
with the previous conditions on the $\mu_a$, we see that a $p \ne 0$
solution exists in the interior region if
\begin{equation}
    \mu_\perp > \mu_\parallel > -{ F\over 4} \, , \qquad\qquad p \ne 0\ .
\label{pnonzeromus}
\end{equation}

Equations~(\ref{pzeromus}) and (\ref{pnonzeromus}), together with the
condition $F>0$, are the fundamental conditions that must be satisfied
at the extremal horizon.  To explore the various possibilities, we
must apply these in turn to the stationary points of $U$ that were
enumerated in Sec.~II.  The last two cases can be immediately
eliminated: Case~3 is excluded because both $\mu_a <0$, thus ruling
out the possibility of a $p=0$ exterior solution, while Case~4 has
$F<0$.  This leaves Cases~1 and~2. The former, $u_*=0$ and $h_*=1$,
corresponds to an extremal horizon in the Coulomb region while the
latter, $u_*=\hat u$ and $h_*=\hat h$, gives a horizon in the monopole
core.\footnote{Since it seems unlikely that the minimum energy
solutions will have either $u$ or $h$ change sign, we assume that the
fields are positive at the horizon.  There can, however, be excited
monopoles in which $u(r)$ goes through a zero; see, e.g.,
Ref.~\cite{orig2}.}  We will study these in detail in the next
section.

\section{Behavior near Coulomb and core region extremal horizons}

In Sec.~IV we showed that there are only two possibilities for the
values of the fields at the extremal horizon.  Although the local
analysis that we used cannot tell whether these are consistent with
the existence of a global solution, our numerical results show that
both types of solution actually occur.  In this section we examine
these more closely.  We also consider the nonsingular near-critical
solutions in which $1/A$ has a related minimum.

\subsection{Coulomb region horizons}

The first possibility, $u_*=0$, $h_*=1$, corresponds to a horizon in
the Coulomb region outside the monopole core.
Equation~(\ref{Uathorizon}) implies that $r_*=r_0$,
the extremal Reissner-Nordstrom value, so $\sigma=F=1$.  The matrix
${\cal M}$ is diagonal, with eigenvalues
\begin{eqnarray} 
   \mu_1 &=& {\cal M}_{11} =  (evr_0)^2 -1 = {{{a}} \over 2} -1  \cr
   \mu_2 &=& {\cal M}_{22} =  4 \lambda v^2 r_0^2 = 2 {{b}} {{a}}\ .
\label{lowmus}
\end{eqnarray}
A solution outside the horizon is given by the extremal
Reissner-Nordstrom metric, with $u=0$ and $h=1$ for all $r_*>r_0$.  We
assume that this is the only exterior solution satisfying the boundary
conditions, and concentrate on the interior solution.

There are several possibilities for the behavior of fields just inside
the horizon:

\noindent {\bf Type I}: The singularity in the matter fields is less
singular than $x^{1/2}$ (i.e., $p=0$).  Equation~(\ref{kformula})
implies that $k=1$, 
so near the horizon
\begin{eqnarray} 
      u &=&  C_u |x|^{\gamma_u} + \alpha_u x + \cdots  \cr
      h &=&  1- C_h |x|^{\gamma_h} - \alpha_h x + \cdots   \cr
      {1\over A} &=& x^2 + \cdots
\label{TypeI}
\end{eqnarray}
where the ellipses represent terms that are determined by the terms shown
explicitly.  Here $\alpha_u$ and $\alpha_h$ denote constants whose values are
fixed by Eq.~(\ref{linearcoeff}). $C_u$ and $C_h$ are not determined
locally, but must instead be adjusted so the boundary conditions at
the origin are satisfied.  The exponents $\gamma_u$ and $\gamma_h$ are
solutions of Eq.~(\ref{gammaeqn}) and, depending on the values of
${{a}}$ and ${{b}}$, may or may not be greater than unity.  The
requirement that these exponents both be greater than 1/2 leads to
Eq.~(\ref{pzeromus}), which implies that
\begin{equation}
     {{a}} > {\rm Max} \, \left[ {7\over 2}\, , \, {3 \over 8 {{b}}}
                \right]\ .
\label{TypeIbound}
\end{equation}

\noindent {\bf Type II}: Matter fields with $|x|^{1/2}$ singularities.  There
are two possibilities here, depending on which field has the
singularity.  In the first (Type IIa),
\begin{eqnarray} 
      u &=& p |x|^{1/2}  + C_u |x|^{\gamma_u} + \alpha_u x + \cdots  \cr
      h &=& 1 -  C_h |x|^{\gamma_h} -  \alpha_h x + \cdots   \cr
      {1\over A} &=& kx^2 + \cdots
\end{eqnarray}
while in the second (Type IIb),
\begin{eqnarray} 
      u &=&  C_u |x|^{\gamma_u} + \alpha_u x + \cdots  \cr
      h &=& 1-  p\sqrt{{{a}}} |x|^{1/2} - C_h
              |x|^{\gamma_h} - \alpha_h x + \cdots   \cr
      {1\over A} &=& k x^2 + \cdots
\end{eqnarray}
Here
\begin{equation}
    k = (1 -  p^2/2)^{-2}
\label{kforCoul}
\end{equation}
and
\begin{equation} 
    p^2 = {1\over 2  \mu_\parallel} \left[   4 \mu_\parallel  -1
      \pm   \sqrt{ 1 + 4 \mu_\parallel } \right]
\label{pforCoul}
\end{equation}
with $\mu_\parallel$ equal to the lesser of the
two eigenvalues in Eq.~(\ref{lowmus}).  If $2<{{a}}<7/2$, the 
upper sign must be used in Eq.~(\ref{pforCoul}).
For Type IIa solutions Eq.~(\ref{pnonzeromus}) requires that
either \begin{equation}
      {3\over 2} < {{a}} < {2\over 1-4{{b}}} \qquad {\rm and } \quad
           {{b}} < {1\over 4}
\label{TypeIIaboundone}
\end{equation}
or
\begin{equation}
        {{a}} > {3\over 2}\qquad {\rm and } \quad
           {{b}} > {1\over 4}\ .
\label{TypeIIaboundtwo}
\end{equation}
The corresponding requirement for Type IIb is that
\begin{equation} 
    {{a}} > {2\over 1-4{{b}}} >2 \qquad {\rm and } \quad
           {{b}} < {1\over 4}\ .
\label{TypeIIaboundthree}
\end{equation}
At the boundary between the IIa and IIb regimes, ${{a}} =
2/(1-4{{b}})$, both $\gamma_u$ and $\gamma_h$ are equal to 1/2 and the two
types of solutions merge.
Note that none of these solutions is possible if  ${{a}} < 3/2$; this
point will be important when we compare this analysis to the 
numerical results.

The most important difference between the Type I and the Type II
solutions is in the behavior of the quantity $\sqrt{AB}$.  If $p \ne
0$, one of the matter fields varies as $|x|^{1/2}$ near the horizon.
As a result, the integral in the exponent in Eq.~(\ref{Bsolution})
diverges for any $r<r_0$, and $\sqrt{AB}$ becomes a step function
centered at the horizon.  Horowitz and Ross \cite{naked} have shown
that a particle in geodesic motion across a horizon feels a tidal
force proportional to the logarithmic derivative of $\sqrt{AB}$, and
have used the term ``naked black holes'' to describe certain
near-extremal black holes solutions for which this quantity is large.
Not only do our Coulomb region extremal solutions fit in this
category, but so do the nearby nonsingular solutions.

\subsection{Horizons in the monopole core}

The second possibility allowed by the analysis of Sec.~IV is that
$u_*=\hat u(r_*)$ and $h_*=\hat h(r_*)$, with $\hat u$ and $\hat h$
given by Eqs.~(\ref{uhateq}) and (\ref{hhat}).  This corresponds to a
horizon in the monopole core region.  Unlike the case of a
Coulomb region horizon, there must be a nontrivial exterior solution,
thus giving an extremal black hole with Higgs and gauge boson hair.
Because our numerical results show solutions of this type only for
relatively large values of ${{b}}$, we will find it convenient to use
large-${{b}}$ expansions to simplify some of the algebra below.

Substituting the expressions for $\hat u$ and $\hat h$ into the
potential $U$ gives 
\begin{equation}
     8\pi G r^2 U = {{{a}} \over 2({{b}}-1)} (evr)^{-2} 
     \left[ -1 + 2 {{b}} (evr)^2 - {{b}} (evr)^4 \right]\ .
\end{equation}
Equation~(\ref{Uathorizon}) requires that this be equal to unity at
the horizon. By making use of the fact that $(evr_*)^2=
{{a}}/2\sigma$, this constraint can be rewritten as
\begin{equation}
     {{b}} = {4\sigma (1-\sigma) \over {{a}}^2 -4 {{a}}\sigma + 4 \sigma}\ .
\end{equation}
The roots of this equation determine ${{a}}$ in terms of ${{b}}$ and
$\sigma$.  For large ${{b}}$, the ${{b}}$-dependence is negligible, and 
\begin{equation}
   {{a}} = 2 \sigma\left[ 1 \mp \sqrt{1 - \sigma^{-1}} \right] + O(1/{{b}})\ .
\label{alphasigma}
\end{equation}

In contrast with the previous case, the 
matrix ${\cal M}$ is not diagonal.  Using the equations obeyed by $\hat u$
and $\hat h$, it can be written as 
\begin{eqnarray}
  {\cal M} &=&  \left( \matrix{ 2{\hat u}^2     &   2\sqrt{2} evr_*\,
         \hat u \hat h   
         \cr\cr    
   2\sqrt{2} evr_*\, \hat u \hat h & 4{{b}} e^2v^2r^2_*\, {\hat h}^2 }
      \right)   \cr &{}&\cr
   &=&  \left( \matrix{ 2(1-y^2)  &  2\sqrt{2} y \sqrt{1-y^2}  \cr\cr
     2\sqrt{2} y \sqrt{1-y^2} & 4 {{b}} y^2 - 4(1-y^2) }\right) 
      + O(1/{{b}})
\end{eqnarray}
where in the second line we have defined
\begin{equation}
     y^2 = (evr_*)^2 = {{{a}} \over 2 \sigma}\ .
\end{equation}
The eigenvalues of ${\cal M}$ are 
\begin{eqnarray}
     \mu_1 &=& {2{{b}} {{a}} \over \sigma} + O(1)  \cr \cr
     \mu_2 &=& 2 - {{{a}} \over \sigma} + O(1/{{b}})\ .
\end{eqnarray}

The exterior solution must be of the $p=0$ type, implying that these
eigenvalues must obey Eq.~(\ref{pzeromus}).  This requires that they both
be greater than $3F/4$, where
\begin{eqnarray}
     F &=& {1 \over {{b}} -1}\left[{{{a}}^2{{b}} \over 4 \sigma} -
              \sigma \right] 
              \cr
        &=& {{{a}}^2 \over 4 \sigma} + O(1/{{b}})\ .
\end{eqnarray}  
In the large ${{b}}$ limit,
$\mu_1$ clearly satisfies this condition, while for $\mu_2$ we obtain
the constraint
\begin{equation}
    \sigma > {3\over 32} {{a}}^2 + {1 \over 2} {{a}} + O(1/{{b}})\ .
\label{alphasigma2}
\end{equation}
We also obtain the requirement ${{b}}>1$, since otherwise $M$ has a
negative eigenvalue.
Since ${{a}}$ and $\sigma$ are also related by Eq.~(\ref{alphasigma}),
we can obtain a constraint that depends only on $\sigma$.  To
satisfy both Eqs.~(\ref{alphasigma}) and~(\ref{alphasigma2}),
we must take the upper sign in
Eq.~(\ref{alphasigma}).  We can then combine the two conditions to obtain
\begin{equation}
    6\sigma^2 -{41\over 15} \sigma -4 + O(1/{{b}}) > 0
\end{equation}
which implies that $\sigma > 1.06+ O(1/{{b}})$.  Note in particular
that $\sigma=1$ (i.e., $r_*=r_0$) is not possible for any value of ${{b}}$.

Once this condition is satisfied, there are no further constraints imposed by
local analysis at the horizon.  In particular, the interior solution can be
either of the $p=0$ or the $p \ne 0$ type.  Because ${\cal M}$ is not
diagonal, as it was in the Coulomb case, the same irrational powers
appear in both $u$ and $h$.  Thus, if $p=0$, the fields near the
horizon behave as 
\begin{eqnarray} 
      u &=& \hat u(r_*) + q_1^u C_1 |x|^{\gamma_1} +q_2^u C_2 |x|^{\gamma_2} +
      \alpha_u x + \cdots  \cr 
      h &=& \hat h(r_*) + q_1^h C_1 |x|^{\gamma_1}+ q_2^h C_2 |x|^{\gamma_2} +
      \alpha_h x + \cdots   \cr 
      {1\over A} &=& F x^2 + \cdots
\label{pzerooutside}
\end{eqnarray}
Here $\gamma_1$ and $\gamma_2$ are solutions of Eq.~(\ref{gammaeqn})
corresponding to the two $\mu_a$; as before, their values may be
either greater or less than unity, depending on the values of ${{a}}$
and ${{b}}$.  The constants $q_a^u$ and $q_a^h$ are determined by the
eigenvectors $\theta_a$.  
The values of $\alpha_u$ and $\alpha_h$ are fixed and are the same on both sides of the
horizon, while $C_1$ and $C_2$ can be varied independently inside and outside
the horizon so that the boundary conditions can be satisfied both at the origin
and at spatial infinity.  

If instead $p \ne 0$, the fields outside the
horizon are as in Eq.~(\ref{pzerooutside}), but inside the horizon
\begin{eqnarray} 
      u &=& \hat u(r_*) + p q_2^u |x|^{1/2} +q_1^u C_1 |x|^{\gamma_1}
            +q_2^u C_2 |x|^{\gamma_2}  + \alpha_u x 
            + \cdots  \cr
      h &=& \hat h(r_*) + p q_2^h |x|^{1/2} + q_1^h C_1 |x|^{\gamma_1}+
            q_2^h C_2 |x|^{\gamma_2}  + \alpha_h x 
            + \cdots   \cr
      {1\over A} &=& k x^2 + \cdots
\end{eqnarray}
We have assumed that $\mu_2 < \mu_1$; if the opposite were the case,
the $|x|^{1/2}$ terms would involve $q_1^u$ and $q_1^h$.  Here $k$
and $p$ are given by Eqs.~(\ref{kformula}) and (\ref{pformula}), with
there being two acceptable solutions. As before, the values of
$\alpha_u$ and $\alpha_h$ are determined, although they are no longer the same
as in the external solution.

\section{Comparison with numerical results}

Let us now see how well these analytic results agree with our
numerical results.  We begin with the low-${{b}}$ regime, where the
nonsingular monopole solutions tend toward a critical solution with a
Coulomb region horizon at the Reissner-Nordstrom value $r=r_0$.
Throughout this regime, the critical solutions we find appear to be
the more singular Type II solutions rather than the Type I solutions
of Eq.~(\ref{TypeI}).  This is consistent with our finding (see
Fig.~\ref{fig:crit}) that the values of ${{a_{\rm cr}}}$ are always
less than the lower bound for Type I solutions given in
Eq.~(\ref{TypeIbound}).

Given the values for ${{a_{\rm cr}}}$ as a function of ${{b}}$,
Eqs.~(\ref{TypeIIaboundone}) and (\ref{TypeIIaboundthree}) predict
that there should be a transition at ${{b}} \approx 0.1$ between the
Type IIa regime, where $u \sim |x|^{1/2}$ and the Type IIb regime,
where $1-h$ has the $|x|^{1/2}$ singularity.  At ${{b}}=0.1$, the
lowest ${{b}}$ for which we have a sequence of stable configurations
that approaches a critical black hole solution, we find that the
powers of $u$ and $1-h$ are both close to $1/2$, as predicted for this
transition.  It is curious to note that this is close to the value of
${{b}}$ below which $(1/A)_{\rm min}$ can be a double-valued function
${{a}}$.

Examining the Coulomb-type black hole solutions in more detail, we
find, for ${{b}} \simle 20$, that as $r$ approaches $r_0$ from below,
the values of $d^2(A^{-1})/dx^2$ and $u^2/x$ tend toward the values of
$k$ and $p$ predicted by Eqs.~(\ref{kforCoul}) and (\ref{pforCoul})
with the upper choice of sign.

We observed in Sec.~IIIB that the matter fields displayed exponential
falloffs when written as functions of the proper length $l(r)$ for
low-${{b}}$ type solutions.  To understand this, we note that the field
equation for $u$, Eq.~(\ref{ueqn}), can be written as
\begin{equation}
    {d^2 u \over dl^2} +\left[{d (A^{-1/2})\over dr} + {8\pi G r
      K\over \sqrt{A}} \right] {du\over dl} ={u(u^2-1)\over r^2} +
      e^2v^2 uh^2\ .
\end{equation}
Our analytic results indicate that the quantity in square brackets is
of order unity as $x \rightarrow 0$ for the interior portion of the
extremal solution.  This suggests that this quantity should be a
relatively slowly varying function for large $l$ and hence that the
falloff of $u$ should behave as a decaying exponential with a
slowly-varying decay constant.  As we saw in Fig.~\ref{fig:near}, this
is indeed the case both for the critical and the near-critical
solutions.  A similar analysis can be applied to $1-h$.

From this point of view, the vanishing of $u$ and $1-h$ at the
extremal horizon is a simple consequence of the fact that $l(r_0)
=\infty$.  This analysis also tells us how these fields should behave
as the critical solution is approached.  We concentrate here on $u$,
and restrict ourselves to the range of ${{b}}$ where $u \sim
|x|^{1/2}$ in the critical solution.  Near $\bar r$, the minimum of
$1/A$, we can write\footnote{Note that $c_1^2$ cannot be the same as
the coefficient $k$ that appears in the critical solution, since the
second derivative of $1/A$ has a discontinuity at the horizon of the
critical solution, but is continuous when $\Delta \ne 0$.}
\begin{equation} {1\over A} \approx c_1^2 (r- \bar r)^2 + \Delta^2\ .
\end{equation}
Substituting this into Eq.~(\ref{properdist}), we find that 
\begin{equation}
    \bar l \equiv l(\bar r) \sim c_1^{-1} \ln (1/\Delta)\ .
\end{equation}   
Writing $u$ as a decreasing exponential, $u \sim u_0 \exp(-c_2 l)$, we
then have
\begin{equation}
   u(\bar r) \sim u_0 (1/A)_{\rm min}^q
\end{equation}
with $q=c_2/2c_1$.  This analysis does not give predictions for $c_1$
and $c_2$.  However, we can get an indication of what to expect for
$q$ by recalling that for the critical ($\Delta=0$) case $u \sim
(1/A)^{1/4}$ as the horizon is approached from the inside.  If the
above expressions were exact and $c_1$ and $c_2$ were truly constant
and independent of $\Delta$, this would imply that $q=1/4$.  Turning to
our numerical results, we do indeed see a power law behavior for $u$, with $q$
ranging between 0.23 and 0.28.

A similar analysis can be applied to the jump in $\sqrt{AB}$ near the
minimum of $1/A$.  For the parameter ranges where we find extremal
solutions with $p \ne 0$, this predicts that $\sqrt{AB}_{\rm r=0}$
varies as $(1/A)_{\rm min}^{-p^2/4}$ assuming a normalization of
$\sqrt{AB}_{\rm r=\infty} = 1$.  Not only does this account for the
power law behavior noted in Sec.~IIIB, but the predictions for the
power are borne out by the data within the numerical accuracy, where
the power $p^2/4$ ranges from about 0.7 to unity.

Near ${{b}} = 20$, ${{a}}_{\rm cr}$ equals $3/2$, the lower bound
allowed by the analysis of the previous section for Coulomb-type
solution.  Nevertheless, the qualitative nature of the approach to
criticality does not change as ${{b}}$ increases past this value and
as ${{a}}_{\rm cr}$ becomes smaller than $3/2$.  Indeed, the
transition to core-type horizons does not appear to occur until about
${{b}}= 40$, where ${{a}}_{\rm cr} = 1.4187$.  It does not seem as if
this discrepancy can be attributed simply to numerical errors.  We
find that changing numerical parameters or modifying details of the
algorithm gives a variation in the value of $a_{\rm cr}$ at the
transition that is at most of order 0.01.  For more details of the
numerical aspect of this issue, see the Appendix.

The resolution of this puzzle seems to lie beyond the precision of our
numerical simulations.  Several possibilities suggest themselves.
First, there may be a sharp
change in the approach to criticality that is only seen when one reaches
solutions for which $(1/A)_{\rm min}$ is less than the limits set by
our spatial step size, in such a way that our extrapolation to the
critical limit is incorrect.  It could be that the decrease in
the minimum of $1/A$ suddenly slows, and that it only reaches zero at 
${{a}}_{\rm cr} > 3/2$, or it could be that a new family of solutions
intervenes.  
Alternatively, it may be that ${{a}}_{\rm cr}$ is correctly
extrapolated, but that the critical solution violates one or more of
the assumptions that we enumerated at the beginning of Sec.~IV.  For
example, the critical solution might have discontinuities in the
matter fields at the horizon; because this would occur at a zero in
$1/A$, it would not necessarily lead to a divergent energy.

We turn now to the high-${{b}}$ regime, where the critical solution
has an extremal horizon in the core region, with $u_*=\hat u(r_*)$,
$h_*=\hat h(r_*)$.  The numerical solutions all have $p=0$ for both
the internal and external solutions, with the matter fields varying
less rapidly than $|x|^{1/2}$ near the horizon.  Moreover, for the
entire region where we observe this type of black hole solution,
the leading behavior of $u-u_*$ and $h-h_*$ at the horizon appears to
be linear in $x$, with the nonanalytic terms being subdominant.
The relations between ${{a}}$, $u_*$, $h_*$ and $r_*$ are all in
agreement with our analytic results.  

Examining the solutions in more detail, we find that in most of the
interior portion of the critical and near-critical solutions both
   $u(r)$ and $h(r)$ closely track $\hat u(r)$ and $\hat h(r)$.  This is
easily understood by recalling the form of the matter action,
Eq.~(\ref{matteraction}).  Once $1/A$ has become small, the gradient
terms in the action are much less important than the
position-dependent potential $U$.  As a result, the fields that
minimize the action should point-by-point be close to the minimum of
$U$, namely $\hat u(r)$ and $\hat h(r)$.

It may be at first surprising that a second minimum in $1/A$ exists at
$r\approx r_0$ outside the true horizon.  To explain this, consider
the behavior of the metric outside the monopole core. If the matter
fields $u$ and $h$ are negligibly small for $r> R_{\rm core}$, then in
this Coulomb region the mass function for a configuration with
magnetic charge $4\pi /e$ can be approximated by
\begin{equation}
    m(r) = M - {2\pi \over e^2}\ .
\end{equation}
This gives a local maximum of $m(r)/r$, corresponding to a local
minimum of $1/A$, at
\begin{equation}
     r_{\rm min} = {4\pi \over e^2 M} = r_0 \left( {M_0\over M}
     \right)
\end{equation}
with the value of $1/A$ at this minimum being $1-(M/M_0)^2$.  This
minimum will occur, regardless of whether or not there is a horizon at
a smaller value of $r$, as long as $r_{\rm min} > R_{\rm core}$.  It
is absent in our solutions for very large ${{b}}$ because for those
solutions ${{a}}$, and thus $v$, are small enough that the core
extends beyond $r_0$.

\section{Concluding Remarks}

We have used both analytic and numerical methods to study
self-gravitating Yang-Mills-Higgs magnetic monopoles for a range of
parameters, with an emphasis on their approach to the black hole
threshold as the Higgs expectation value tends toward its critical
value.  We find two quite distinct behaviors.  In the low-$b$ regime,
with weak Higgs self-coupling, the critical solution is identical to
the extremal Reissner-Nordstrom solution outside the horizon.  Inside
the horizon, the matter fields are smooth and nonsingular, as is
$1/A$.  However, because of the step function behavior of $\sqrt{AB}$
at the horizon, $B=g_{tt}$ vanishes identically in the interior.  The
associated singularity at the horizon means that an observer falling
freely into a critical or near-critical monopole will experience the
strong tidal forces characteristic of a Horowitz-Ross naked black
hole.

One might speculate that these singularities were inextricably
associated with the transition from a nonsingular monopole to a black
hole spacetime and were independent of the details of the Higgs
potential.  However, our solutions in the large-$b$ regime show that
this is not the case.  For these, the critical solution has nontrivial
matter fields outside the horizon, and thus is an extremal black hole
with hair.  Although the metric functions are still nonanalytic at the
horizon, their singularities are much weaker and do not lead to large
tidal forces.

The transition between these two regimes is itself quite interesting,
and is in many ways reminiscent of a first-order transition.  For a
range of values of $b$ on both sides of the transition point, $1/A$
for the critical solution has two distinct minima.  For values of $b$
below the transition, $1/A$ is positive at the inner minimum and has a
double zero at the outer minimum at $r=r_0$.  The inner minimum moves
downward with increasing $b$ until, at the transition point, there are
two distinct but degenerate minima, with $1/A$ vanishing at both.  As
$b$ is increased further, the outer minimum moves upward while the
inner minimum remains a zero of $1/A$.  While this implies a
discontinuity in the horizon area, the mass of the black hole is
continuous.  In terms of black hole thermodynamics, this corresponds
to a discontinuity in the entropy but (because these are zero
temperature black holes) a free energy that is a continuous function of
$b$.

Our focus in this paper has been on the behavior of the monopole
solutions as the Higgs expectation value is increased to its critical
value.  It seems natural to ask how these solutions evolve if $v$ is
increased still further.  Actually, there is a bifurcation at this
point.  If one requires that the solutions remain well-behaved at
spatial infinity, the solutions with $v > v_{\rm cr}$ are black hole
solutions with singularities at $r=0$.  In the low-$b$ regime, these
are simply nonextremal Reissner-Nordstrom black holes, of varying
mass, with no dependence on either $a$ or $b$.  (There are also
magnetically-charged black holes with hair in this regime; however,
they are not continuously connected to our critical solutions.)  In
the high-$b$ regime, the counting of boundary conditions suggests
that, for given values of $a$ and $b$, the continuation of the
external solution is a one-parameter family of black holes with hair.
Alternatively, one can require that the fields remain nonsingular at
$r=0$ and continue the interior solution.  Because there are three
boundary conditions at the origin, compared to the two at spatial
infinity, this gives solutions with nonsingular interiors bounded by
horizons that are uniquely determined by specifying $a$ and $b$.  The
behavior of the metric near the horizon is similar to that found when
de Sitter spacetime is written in static coordinates, suggesting a
cosmological interpretation for these solutions.  We suspect that,
possibly with a suitably modified Higgs potential, these may give rise
to topological inflation \cite{infl,infl2}.

Finally, having studied these self-gravitating monopoles and obtained
an understanding of their approach to criticality, we can use these
solutions as tools for investigating the transition from a nonsingular
spacetime to one containing blackhole horizons.  We will describe this
elsewhere.

\acknowledgements

We wish to thank Dieter Maison, Gary Horowitz, and Mark Trodden
for useful conversations.  This work was supported in part by the
U.S. Department of Energy.

\pagebreak

\appendix
\section{Numerical Discussion of Transition Regime}

In Section~VI we noted that a discrepancy appears between the
analytic argument that low-$b$ type critical solutions can only exist
when $a_{\rm cr} > 1.5$ and the numerical observation that this type
of critical solution persists when $a_{\rm cr} < 1.5$.  We discuss
here the details of our investigation that lead us to believe that
this discrepancy is not a numerical artifact.

Recall that the transition between the high and low-$b$ type critical
solutions is reminiscent of a first-order phase transition.  The
metric function $1/A(r)$ for a regular monopole solution near
criticality with $b$ near the transition value possesses two minima,
one near the extremal Reissner-Nordstrom radius, $r_0$, the other in
the core of the monopole, $r < r_0$.  As $a$ is increased (for a fixed $b$),
one of these minima becomes a horizon, and a critical solution is
attained.  If $1/A$ at the inner minimum becomes zero first, then we
have a high-$b$ type critical solution; when $1/A$ for the outer
minimum touches down first, we have low-$b$ type solution.  Let us
examine each of these minima separately for regular solutions
near criticality in this transition region.

\subsection{The core minimum}

The inner minimum for the types of solutions under consideration
varies rapidly with both $a$ and $b$, so one can identify very
precisely when such a minimum becomes a horizon.  Moreover, the field
variables and their derivatives are well-behaved near this radius, and
the numerical solutions here are robust and insensitive to variations
in the numerical algorithm.  Different numerical implementations of
the field equations leads to differences in $a_{\rm cr}$ of a few
parts in $10^4$ for a high-$b$ type critical solution.

The numerical and analytic results could be reconciled if (1) the 
high-$b$ type solutions persisted down to lower $b$ into a region  where
$a_{\rm cr} > 1.5$ and (2) the identification of the outer minimum   as a
zero of $1/A$ for the critical solutions in this region was a result of
numerical error.   We can address the first issue by using a trick. In
Sec.~IV  we learned that critical solutions solve separate interior and
exterior boundary value problems.  Hence, we can determine  $a_{\rm cr}$
as a function of $b$ for the high-$b$ type solutions by just solving the
field equations from the origin to the horizon, ignoring the exterior
region from the horizon out to  spatial infinity.   By using an algorithm
similar to that described in Sec.~III, we find  the family of critical
solutions represented by the dashed line in the inset  of
Figure~\ref{fig:crit}.  These solutions have high-$b$ type horizons but
are not asymptotically flat at spatial infinity; however, if there were
an asymptotically flat critical solution with the same value of $b$, it
would have the same $a_{\rm cr}$ as these solutions. This dashed line
crosses $a_{\rm cr} = 1.5$ when $b = 25.6$.  Thus to be consistent with
the  analytic results, the transition value between low and high-$b$
critical solutions {\em must} occur for $b < 25.6$ (rather that at  the
value of $b = 40$ indicated by our numerical results).  Again, this result
appears to be numerically robust to a few parts in $10^4$.

\subsection{The Coulomb minimum}

According to the previous discussion, the ability to reconcile
numerical and analytic results now depends on the sensitivity of the
outer minimum, the one near $r_0$, to numerical error.  From the inset
in Figure~\ref{fig:crit}, we see that in order for $b_{\rm tr} <
25.6$, our values for $a_{\rm cr}$ for the low-$b$ type critical solutions must
be in error (too low) by approximately $0.04$.  A priori, such an error is
certainly plausible. At this minimum, the solutions near criticality vary
sluggishly with $a$ and $b$.  The spatial derivatives of field variables are
large at this radius, and indeed are expected to be singular if the outer
minimum becomes a horizon.  It is here that we are most susceptible to
numerical errors.

As we discussed in Sec.~III, discretized versions of
Eqs.~(\ref{ueqn})--(\ref{Meqn}) are implemented in our numerical
algorithm to identify static monopole solutions.  In order to probe
the robustness of the algorithm and the confidence with which we can
treat these solutions, different implementations of the discretization
were examined and compared.  In particular, we focus on
Eqs.~(\ref{ueqn}) and (\ref{heqn}).  First, we used the straightforward
expansion of these equations.
\begin{eqnarray}
	{1 \over A}{d^2u \over dr^2}
	&+& {du \over dr}\left[{d\over dr}\left({1\over A}\right)
		+ 8\pi G {rK\over A}\right] = \cdots
	\label{A1}	\\
	{1 \over A}{d^2h \over dr^2}
	&+& {dh \over dr}\left[{d\over dr}\left({1\over A}\right)
		+ 8\pi G {rK\over A} + {2r\over A}\right] = \cdots\ .
	\nonumber
\end{eqnarray}
In a separate treatment, we used Eq.~(\ref{Meqn}) to eliminate
$d(1/A)/dr$.
\begin{eqnarray}
	{1 \over A}{d^2u \over dr^2}
	+ {1\over r}{du \over dr}
	\left[(1 - 8\pi G r^2U) - {1\over A}\right] &=& \cdots
	\label{A2}		\\
	{1 \over A}{d^2h \over dr^2}
	+ {1\over r}{dh \over dr}
	\left[(1 - 8\pi G r^2U) + {1\over A}\right] &=& \cdots\ .
	\nonumber
\end{eqnarray}
Though these two systems of equations are equivalent in the continuum
limit, their discretized forms which appear in the numerical
implementation suffer a ${\cal O}[(\Delta r)^2]$ difference.   

\begin{figure} \begin{center}\PSbox{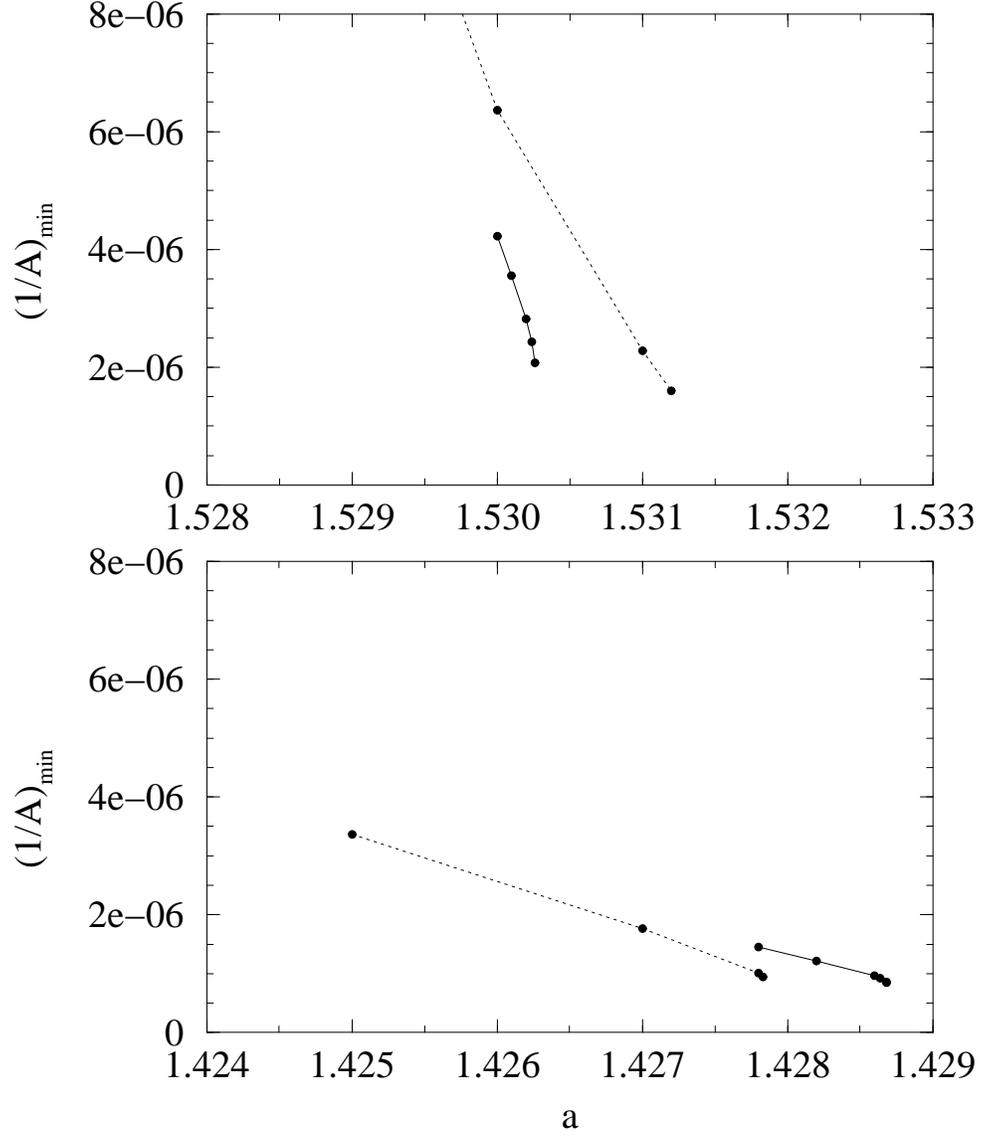
hscale=100 vscale=100 hoffset=-120 voffset=-35}{5in}{6in}\end{center}
\caption{Comparison of $(1/A)_{\rm min}$ versus $a$ using two
different discretizations at $b=16$ (top diagram) and $b=35$
(bottom diagram).  The dotted curve employs the system Eqs.~(A1)
while the solid curve employs the system Eqs.~(A2).} 
\label{fig:app}
\end{figure}

Figure~\ref{fig:app} shows the $(1/A)_{\rm min}$ as a function of $a$
for solutions near criticality using the two different equation sets
Eq.~(\ref{A1}) and (\ref{A2}) at $b=16$ and $b=35$.  The curves
in Fig.~\ref{fig:app} do not extend into the region where $(1/A)_{\rm
min} \lesssim 1.0\times 10^{-6}$ since this is ${\cal O}[(\Delta
r)^2]$ and our solutions break down.  (For these calculations we used $\Delta
r= 5 \times 10^{-4}$.)  Indeed we find no stable solutions at all in this
regime;  however, we expect this effect is an artifact of the discretization. 
What happens for this very small $(1/A)_{\rm min}$ is beyond the ability of our
resources to probe.

Nevertheless, the extrapolated difference in $a_{\rm cr}$ between the
two numerical implementations were on the order $0.001$.  Other similar
variations in the numerical algorithm were tried, and analogous
results are found.  Moreover, varying the spatial discretization size
also yielded differences in $a_{\rm cr}$ on the order $0.001$.  Thus,
numerical error does not seem to provide the $\Delta a_{\rm cr} \sim
0.04$ needed to account for the discrepancy between the numerical
observations and the analytic results.  To illustrate the scope of the
discrepancy needed, the difference between the two curves needs to be
approximately eight times the total range of the horizontal axis shown
in Fig.~\ref{fig:app}.

Finally, there is no apparent qualitative change between solutions whose
critical $a_{\rm cr}$ is greater than or less than 1.5, as seen in
Fig.~\ref{fig:app}.  The approach of
$(1/A)_{\rm min}$ to zero is a much slower function of $a$ as $b$
increases, but there is no sharp change at $a_{\rm cr}=1.5$.


\begin{thebibliography}{}


\bibitem{orig1}
K. Lee, V. P. Nair, and E. J. Weinberg, Phys. Rev. D
{\bf 45}, 2751 (1992).

\bibitem{ortiz}
M. E. Ortiz, \pr{45}{R2586}92.

\bibitem{orig2}
P. Breitenlohner, P. Forg\'{a}cs, and D. Maison,
Nucl. Phys. {\bf B383}, 357 (1992);

\bibitem{orig2b}
P. Breitenlohner, P. Forg\'{a}cs, and D. Maison,
Nucl. Phys. {\bf B442}, 126 (1995).

\bibitem{AB}
P. C. Aichelburg and P. Bizon, Phys. Rev. D {\bf 48}, 607 (1993).

\bibitem{review} See M. S. Volkov and D. V. Gal'tsov, hep-th/9810070,
and references therein; For more recent work, see Y. Brihaye,
B. Hartmann, J. Kunz and N. Tell, hep-th/9904065; P. K. Tripathy,
hep-th/9904186; P. K. Tripathy, hep-th/9906164; T. Tamaki, K. Maeda
and T. Torii, gr-qc/9906099; S. L. Liebling, gr-qc/9906014; N. Grandi,
R. L . Pakman, F. A.  Schaposnik, and G. Silva, hep-th/9906244.

\bibitem{future}
A. Lue and E. J. Weinberg, in preparation.

\bibitem{vanN}
P. van Nieuwenhuizen, D. Wilkinson, and M. J. Perry, \pr{13}{778}76.

\bibitem{stability}
H. Hollmann, Phys. Lett. B {\bf 338}, 181 (1994).

\bibitem{stability2}
T. Tachizawa, K. Maeda, and T. Torii, \pr{51}{4054}95.

\bibitem{hajicek}
P. Hajicek, Proc. R. Soc. London {\bf A386}, 223 (1983);
J. Phys A {\bf 16}, 1191 (1983).

\bibitem{naked}
G. T. Horowitz and S. F. Ross, \pr{56}{2180}97;
{\em ibid.} {\bf 57}, 1098 (1998).

\bibitem{infl}
A. Vilenkin, Phys. Rev. Lett. {\bf 72}, 3137 (1994).

\bibitem{infl2}
A. Linde, Phys. Lett. B {\bf 327}, 208 (1994).

\end{thebibliography}
\end{document}